\documentclass[twocolumn]{autart}    

\usepackage{graphicx}          
\usepackage{color}
\usepackage{booktabs}
\usepackage{amsmath,amssymb,amsfonts}
\usepackage{algorithm}
\usepackage{subfigure} 
\usepackage[noend]{algpseudocode}
\newcommand{\be}{\begin{equation}}
	\newcommand{\ee}{\end{equation}}
\newcommand{\bea}{\begin{eqnarray}}
	\newcommand{\eea}{\end{eqnarray}}
\def\proof{\noindent{\bf{Proof.} }}
\begin{document}

\begin{frontmatter}
\title{Infinite-Horizon Optimal Wireless Control Over Shared State-Dependent Fading Channels for IIoT Systems\thanksref{footnoteinfo}} 

\thanks[footnoteinfo]{The work was supported by National Key
	R\&D Program of China under the grant 2022YFB3303900 and
	the NSF of China under the grants 61933009, 62103272. This paper was not presented at any IFAC  meeting. Corresponding author S.~Zhu.}

\author{Shuling~Wang}\ead{shulingwang2021\_sjtu@sjtu.edu.cn},    
\author{Peizhe~Li}\ead{lipeizhe2020@sjtu.edu.cn},               
\author{Shanying~Zhu}\ead{shyzhu@sjtu.edu.cn},  
\author{Cailian~Chen}\ead{cailianchen@sjtu.edu.cn},  
\author{Xinping~Guan}\ead{xpguan@sjtu.edu.cn}  
\address{Department of Automation, Shanghai Jiao Tong University,
	Shanghai 200240, China;
	Key Laboratory of System Control and Information Processing, Ministry of Education of China,
	Shanghai 200240, China, and also Shanghai Engineering Research Center of Intelligent Control
	and Management, Shanghai 200240, China}  


\begin{abstract}                          
Heterogeneous systems consisting of a multiloop wireless control system (WCS) and a
mobile agent system (MAS) are ubiquitous in Industrial Internet of Things systems.
Within these systems, the positions of mobile agents may lead to shadow
fading on the wireless channel that the WCS is controlled over and can significantly
compromise its performance, requiring joint coordination between the WCS and MAS. 
Such coordination introduces different time steps  and hybrid state spaces consisting of  logical components and  continuous components. This paper focuses on the infinite-horizon optimal control of MAS to ensure the performance of WCS while minimizing an average cost for the heterogeneous system subject to safety constraints. A state-dependent fading channel is  modeled to capture  
interference among transmission links, as well as 
the effects of mobile agents' movements on  successful wireless transmission. 
In order to address the heterogeneous system dynamics, the optimal control  problem is formulated as the optimal constrained set stabilization of the MAS by establishing a necessary and sufficient condition for the Lyapunov-like performance of WCS with the expected decay rates. 
Using the semi-tensor product  of matrices, a constrained optimal state transition graph is constructed to encode 
the constrained system dynamics as well as objective function, which further reduces the problem  into a minimum-mean cycle problem for the graph.  
By studying the properties
of the graph,  
the feasibility is proven, 
and an effective algorithm is proposed for the construction of optimal input
sequences.
 An illustrative example is provided to demonstrate effectiveness of the proposed method.
\end{abstract}

\end{frontmatter}

\section{Introduction}
\label{sec:introduction}
\subsection{Background and motivation}
Industrial Internet of Things (IIoT) enables flexible, efficient and sustainable production in many fields, including smart manufacturing, through numerous plants that can be reconfigured based on process requirements \cite{Baumann2021,fadlullah2011}. These plants consist of multiple physical systems 
	with sensing, computation, communication and actuation capabilities, which flexibly interact with one another and human operators \cite{Ahlen2019,vitturi2019}. Due to such complex interactions among the entities, wired communication will reach its limits. Instead, 
	wireless sensors are deployed to monitor industrial plants, which enable much higher flexibility while reducing cable wrapping and limitations on plant operations, forming multiloop wireless control systems (WCSs). 

An integral aspect of IIoT in smart manufacturing is the use of mobile agents, e.g., automated guided vehicles (AGVs) \cite{Pulikottil2021}, 
which can coordinate with WCSs contributing to overall manufacturing processes in various ways, 
e.g., manufacturing systems with heavy machines and cranes \cite{Agrawal2014}, assembly
processes with autonomous assembly arms and forklifts \cite{hu2022,hu2019}. In these scenarios, the state of agents are  subject to certain constraints to ensure production safety such as moving within task areas and velocity limits while avoiding dangerous positions \cite{Baumann2021}.
Moreover, the positions of mobile agents may induce shadow fading on the wireless channel that the WCS is controlled over 
\cite{Agrawal2014,Quevedo2013}. 
On-site measurements in factories reveal  that  the channel fading caused by machinery movements  exhibits a piecewise behavior, i.e., the fading channel gain remains stationary as the machinery moves within a certain area \cite{Ahlen2019}. And each area is associated with a certain level of shadowing effects. 
This corresponds to a high level dynamics of MAS  exhibiting  logical characteristics. Additionally,  the movements of mobile agents among areas after spatial discretization is typically on the time scale of seconds, while the wireless channel transmits packets on the order of milliseconds.  This means that the coordination of the WCS and mobile agent system (MAS) introduces  different time steps and  
hybrid state spaces, consisting of the  continuous component of industrial plants
and the logical component of mobile agents. 
These  distinctive features make the modeling and analysis tools on traditional WCSs fail in addressing the IIoT system considered  in the paper. Due to the complex interactions between the WCS and MAS,  
in order to ensure the control performance of the WCS, it is necessary to  explicitly examine the influence of  mobile agents' positions on the channel states of the WCS, and to design a constrained  motion strategy for agents such that the desired performance requirements are guaranteed for the WCS.

\subsection{Related work}
Since the typical connection density in the IIoT scenario is $10^6/km^2$, transmission scheduling among sensors over the shared limited wireless medium is an important topic in practice \cite{valerio2021}. 
For example, static schedules typically specifying periodic communication sequences were proposed to satisfy given control performance requirements such as stability \cite{Hristu2001}, controllability and
observability \cite{zhang2006} and minimizing linear quadratic objectives \cite{le2011}.
Dynamic schedules deciding access to the shared wireless medium dynamically at each step were also extensively studied \cite{donkers2011,gatsis2015,Molin}.  For a survey on design and optimization for WCSs, please refer to \cite{Park2018}.

As important components of IIoT systems, the MAS and WCS  are typically considered separately in the literature. 
Recently, the finite-field network architecture, which is a special kind of finite-valued systems, has been adopted to model the analysis and operational control for MAS \cite{Pasqualetti2014}. 
The (in)finite horizon optimal control problems for finite-valued systems have received extensive attention \cite{Fornasini2014,wuy2019,Zhao2011}.
It is noted that most existing results are established for the optimal control of finite-valued systems with only system dynamics constraints. 
Taking the state constraints into account,  \cite{gao2021-1} proposed an efficient graph-theoretical approach for switched finite-valued systems. 
Noting that all the above results are established for finite-valued systems, 
i.e., the dynamics of all nodes in the system are the same, they are not applicable for the optimization of heterogeneous IIoT systems subject to
control performance requirements constraints.
 As for the other part of IIoT, WCSs with fading channels have been extensively studied in the past few decades. It is noted that the channel gains used to characterize shadow fading are traditionally modeled as independent identical distributed random processes \cite{Gatsis2014} or Markov chains \cite{zhangq1999}, where the network state is assumed to be independent from
physical states. With such independency, several results were established for WCSs over fading channels including co-design problem for communication
and control systems \cite{Gatsis2014}, state estimation \cite{Quevedo2012} and so on.
References \cite{gatsis2015} and \cite{gatsis2018} considered shared fading channels, where centralized and distributed channel-aware access mechanisms were respectively presented to meet the Lyapunov-like performance requirements while minimizing the total expected power consumption. The above channel models are clearly inadequate for characterizing the complex interactions between the WCS and MAS. In addition, all these results focus on the control performance of plants with only continuous state
space. It is obvious that these methods fail in analyzing IIoT systems
with hybrid state spaces considered in this paper.

Few works addressing dependence of network state on physical states have been
conducted in IIoT. 
A state-dependent bursty fading channel model was proposed in \cite{hu2014}. 
However, only homogeneous systems without constraints were considered in these works. In order to model random time variations of wireless environment,  \cite{Quevedo2013} introduced a network state process for state estimation, but the network topology was assumed to be fixed. The optimal network topology configuration was then determined for each network state to minimize an expected error covariance measure \cite{Leong2016}. However, the external environment in \cite{Leong2016,Quevedo2013} was modeled as a (semi)-Markov chain which cannot be controlled. 
Removing the uncontrollability assumptions and modeling the external environment as a Markov decision process, \cite{hu2019} studied the  infinite-horizon optimal design problem of communication and control, such that the single-loop WCS with state-dependent channel  achieves both safety and efficiency. By formulating  this problem as a constrained two-player cooperative game, approximate optimal control and transmission power policies were obtained by solving relaxed convex generalized geometric programs. Similar problem was studied in \cite{hu2022}, which considered a generalized state-dependent Markov channel model to address time varying data rates. By solving a constrained polynomial optimization problem, optimal design strategies were obtained. In the above works, a common time step is adopted for the evolution of the MAS and WCS. Noting that the wireless channel typically transmit packets on  the order of  milliseconds, this 
requires mobile agents to move   
on a millisecond time scale, which is often unrealistic.  
In addition, for the practical smart manufacturing scenarios where several plants are controlled over a shared wireless medium, 
the state-dependent channel models cannot capture the interference among transmission links, and the coupling between WCS and MAS.

\subsection{Contributions}
In this paper, we 
focus on the optimal control of MAS with safety constraints to ensure the Lyapunov-like performance 
of WCS, while minimizing an infinite-horizon average cost for the IIoT system in the presence of WCS and MAS coupling. 
A preliminary version of the paper appears in \cite{wangs2022}, where a common time step is considered for the WCS and 
	the leader-follower MAS. 
The main contributions of the paper are summarized as follows. 

\begin{itemize}
\item{We propose a heterogeneous IIoT system framework to capture the coupling between the WCS and MAS, where the multiloop
WCS is described as a switched dynamics model depending on wireless transmission, and the finite-field  architecture is  used to model the high level dynamics of MAS moving among  areas with different levels of shadow effects. 
Different time steps are
adopted to capture different time scales between the wireless transmission of control
systems and the  movements of mobile agents among discretized areas. Effect of mobile agents' movements on successful wireless transmission is characterized by a state-dependent fading channel, which captures the spectrum resource competition and interference among different loops. 
Based on this model, the infinite-horizon optimal control problem is established  for the heterogeneous IIoT system to 
quantify the tradeoff between WCS's 
power consumption and MAS's control cost, while ensuring the Lyapunov-like performance requirement of the WCS and the safety constraints of the MAS.}

\item{To address the hybrid nature of the heterogeneous IIoT system, we first equivalently convert the optimal control problem into the  optimal constrained set stabilization problem of the MAS.  
		In order to address the nonlinearity due to the finite-field model of the MAS,   we adopt the semi-tensor product  approach, 
		and construct a constrained optimal state transition graph  
		to handle 
		safety constraints  
		elegantly.  
		By encoding the constrained system dynamics and objective function into the graph,
		the problem  is further deduced  into a minimum-mean cycle problem on the constrained optimal state transition graph. Then, through studying the properties of the graph, we propose an efficient algorithm to design optimal input sequences that ensure wireless control performance and safety constraints.} 
\end{itemize}

This paper is organized as follows. In Section 2, we characterize the heterogeneous IIoT system and formulate the problem studied in this paper. 
Section 3 formulates the optimal control problem as the optimal constrained set stabilization problem for the MAS. 
Design of optimal controllers using the graphical method is provided in Section 4. 
Section 5  demonstrates main results by an illustrative example, which is followed by the conclusion in Section 6. The key notations used in this paper are summarized in Table \ref{table0}. Throughout this paper, semi-tensor product is the basic matrix product defined as $M\ltimes P=(M\otimes I_{l/n})(P\otimes I_{l/p})$, where $M\in{\mathbb{R}}^{m\times n}$, $P\in {\mathbb{R}}^{p\times q}$, $l$ is the least common multiple of $n$ and $p$, and  $\otimes$ is the Kronecker product of matrices \cite{chengqi2011,cheng2011}. Since the semi-tensor product is a generalization of the conventional matrix product, the symbol ``$\ltimes$'' is omitted in most places of this paper when no confusion arises.

\begin{table}[!]
	\caption{Notations.}	\centering
	\begin{tabular}{l l}
		\toprule[0.8pt]
		Notations & ~Definitions\\
		\midrule 
		$\mathcal{D}_s$ &  ~Logic domain $\{0,1,\cdots,s-1\}$ \\
		$I_s$ & ~$s$-dimensional identity matrix \\
		$\delta_s^i$ & ~$i$-th column of $I_s$ \\
		$[F]_{:,j}$ & ~$j$-th column of matrix $F$ \\
		$[F]_{i,j}$  & ~$(i,j)$-th entry of matrix $F$ \\
		$\mathbb{R}^{n\times m}$ & ~Set of $n\times m$ real matrices  \\
		$Col(F)$ & ~Set  $\{[F]_{:,j}:j=1,\cdots,m\}$ for $F\in\mathbb{R}^{n\times m}$\\
		$\ltimes$ & ~Semi-tensor product \\
		$\lfloor\cdot\rfloor$ & ~Floor of a real number \\
		$\mathbb{P}\{\cdot\}$ & ~Probability of a random event\\
		$\mathbb{E}\{\cdot\}$ & ~Expectation of a random variable\\
		$|\mathcal{M}|$ & ~Cardinal number of set $\mathcal{M}$ \\
		\bottomrule [0.8pt]
	\end{tabular}
	\label{table0}
\end{table}



\section{System model and problem formulation}
\begin{figure}[!t]
	\centerline{\includegraphics[width=3.1in]{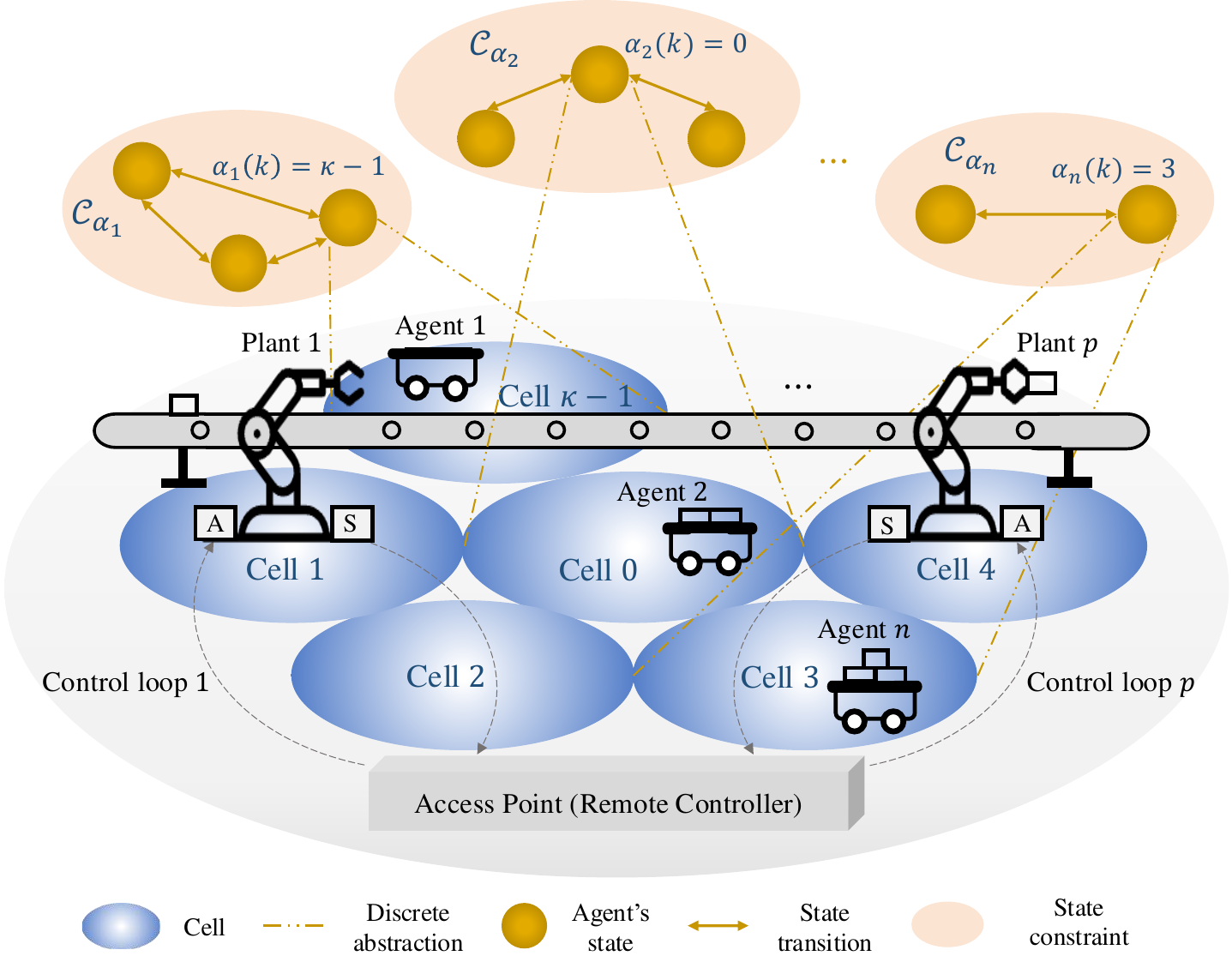}}
	\caption{Framework of a heterogeneous IIoT consisting of a multiloop WCS and an MAS in a factory floor. 
		`S' and `A' respectively
		represent sensing and actuation capabilities. 
The workspace is partitioned into  $\kappa$ cells $\{0,1,\cdots, \kappa-1\}$, and each cell is abstracted as a logical state indicating a level of shadowing effects.  Each mobile agent $j$ moves within its task area, leading to logical state transitions among cells  in the state constraint set $\mathcal{C}_{\alpha_j}$.}
	\label{fig01}
\end{figure}

We consider a heterogeneous IIoT system where a multiloop WCS and an MAS
coordinate with each other to jointly perform overall tasks in a typical smart manufacturing
scenario (Fig. \ref{fig01}).
In the wireless control architecture, $q$ independent plants are controlled over a shared wireless medium, where the measurements of plant $i$, $i\in\{1,\cdots,q\}$ are wirelessly transmitted to the access point by sensor $i$ to compute the
control inputs. At the same time, $n$ mobile agents perform manufacturing tasks in the same workspace. 
In this process, positions
of mobile agents may lead to shadow fading on the wireless channel that the WCS is controlled over, resulting in the coupling between the WCS and MAS.

\subsection{Plants }
Denote the state of plant $i$ at time instant $t_l$ by $x_i(l)\in\mathbb{R}^{n_i}$, $i=1,\cdots,q$.
The discrete-time model of plant $i$ is
\be\label{eq01-1}
x_i(l+1)=A_ix_i(l)+B_iv_i(l)+\xi_i(l),
\ee
where $\{t_l\}_{l=0}^\infty$ with $t_l<t_{l+1}$ is the sequence of sampling time instants, $A_i\in \mathbb{R}^{n_i\times n_i}$ and $B_i\in \mathbb{R}^{n_i\times m_i}$ are system matrix and input matrix, respectively,  $v_i\in\mathbb{R}^{m_i}$ is the driving input, and $\{\xi_i(l):l=0,1,\cdots\}$ represents an independent identically distributed noise process with mean zero and real positive semi-definite covariance $\Xi_i$.

\subsection{Sensors and access point}
$q$ sensors are placed to monitor the plants. 
At every time instant $t_l$, each sensor $i$ takes a measurement and chooses whether transmitting over the shared channel or not. 
We adopt the discrete channel model \cite{Sadeghi2008}, i.e.,  the local channel state
of system $i$ at time instant $t_l$ is $\gamma_i(l)\in\mathcal{D}_r$.
Assume that the wireless communication policy\footnote{
	Regardless of the effect of mobile agents' movements
	on wireless transmission, following the approach in \cite{gatsis2018}, it can be proven that the optimal transmission policies ensuring the Lyapunov-like control performance of the WCS only depend on the channel states $\gamma_i(l)$. 
} is pre-designed as
\be\label{eq04}
\varphi_i(l)=h_i(\gamma_i(l)),
\ee
where $h_i:\mathcal{D}_r\rightarrow\mathcal{D}_2$, $\varphi_i(l)=1$ (or $0$) means that sensor $i$ transmits (or does not transmit) over the shared channel at time instant $t_l$, and each sensor transmits only when its corresponding local channel state is favorable enough and backs off otherwise. 
In practice,  before deciding whether to transmit over the shared channel,  each sensor can acquire the local channel state information by either directly using the visual sensor to observe the positions of the mobile agents, or a short pilot signal sent from the access point to the sensor at the beginning of each time slot. The pilot signal may also serve as a synchronization purpose \cite{gatsis2018,hu2011}.  

When sensor $i$ decides to transmit, it wirelessly transmits measurements of plant $i$ to the access point with a fixed power $\mu_i>0$, where the access point plays the role of remote controller responsible for computing control decisions based on the received measurements \cite{ma2018}. Then, control decisions are wirelessly fed back to actuator $i$. These $q$ 
control loops form a multiloop WCS.   


\subsection{Multiloop WCS}
The transmission of measurements and control decisions might fail due to packet decoding errors and packet collisions.  Assume that both wireless uplink and downlink  have the same channel characteristics \cite{hu2019}. In this case, we focus on the uplink transmission
	from the sensors to the access point, and the idea can be extended into the downlink transmission scenario straightforwardly. 
Use $\lambda_{i}(l)\in\mathcal{D}_2$ to indicate the successful or fail transmission of sensor information for plant $i$ at time instant $t_l$. Then, the measurement $x_i(l)$ is used to compute the control input $v_i(l)$ 
at a successful transmission ($\lambda_{i}(l)=1$), leading to the asymptotically stable closed-loop system matrix $A_{c,i}$. 
Otherwise, an open-loop system with the system matrix $A_{o,i}$, which may be unstable, is obtained. The resulting system can be described as the following switched model:
\be\label{eq01}
x_i(l+1)=\left\{
\begin{array}{ll}
A_{c,i}x_i(l)+\xi_i(l), & \lambda_i(l)=1, \\
A_{o,i}x_i(l)+\xi_i(l), & \lambda_i(l)=0,
\end{array}
\right.~i=1,\cdots,q.
\ee

\subsection{Finite-field MAS}
An MAS coordinates with the multiloop WCS to jointly perform smart manufacturing tasks. 
Mobile agents are driven by the standard single-integrator dynamics\footnote{
This paper uses the model as an illustration of  how to abstract the finite-field  MAS from the  physics dynamics. The proposed results are general and can be extended to other models for mobile agents.} as
$$z_j(k+1)=z_j(k)+T\nu_j(k), j=1,\cdots,n,$$
where $z_j(k),\nu_j(k)\in\mathbb{R}^{2}$ respectively denote the position and velocity of agent $j$ in the two-dimensional space
at  time instant $t_{\tau k}$,  $\{t_{\tau k}\}_{k=0}^\infty$ is the sequence of sampling time instants with time step $T$, and $\tau$ is a positive integer indicating different time steps for the evolution of the MAS and WCS. 

When coordinating with the WCS to jointly perform overall tasks, the positions of
 	mobile agents may lead to shadow fading on the wireless channel that the WCS
 	is controlled over.  Note that what matters for shadow fading  is the position in space within certain
 	intervals \cite{Ahlen2019}, not the specific position determined by the single-integrator dynamics.
Partition the whole workspace into $\kappa$ non-overlapping   cells  $\{0,1,\cdots, \kappa-1\}$. 
 Abstract the cell in which each mobile agent is located   as a logical state in $\mathcal{D}_\kappa$. 
Each agent interacts with others by on-board sensors, and updates its state  as a weighted combination of its previous state, those of its in-neighbors and the input, 
determined by the following finite-field model:
\be\label{eq02}
\alpha_j(k+1)\!=\!\{_\kappa\!\sum\limits_{l\in\mathcal{I}_j\cup\{j\}}\} a_{j,l}\times_{\kappa}\alpha_l(k)
+_{\kappa}u_j(k), j=1,\cdots,n,
\ee
where  $\mathcal{I}_j$ is the set of all in-neighbors of $j$, $a_{j,l}\in\mathcal{D}_\kappa$ denotes the weight, operations $+_\kappa$ and  $\times_\kappa$ respectively denote the modular addition and modular multiplication over $\mathcal{D}_\kappa$ \cite{Lidl1996}, 
$\alpha_j(k)$, $u_j(k)\in\mathcal{D}_\kappa$ respectively denote the 
state and  control input of agent $j$ at time instant $t_{\tau k}$. 
The feedback term $\{_\kappa\sum_{l\in\mathcal{I}_j}\} a_{j,l}\times_{\kappa}\alpha_l(k)$ together with $u_j(k)$ can describe several mobile agents' controllers, including consensus, rendezvous and formation \cite{Mesbahi,Pasqualetti2014,Sundaram2013}. 
The constraint on state $\alpha(k)=(\alpha_1(k),\cdots,\alpha_n(k))\in\mathcal{C}_\alpha\subseteq\mathcal{D}_\kappa\times\cdots\times\mathcal{D}_\kappa$ with $\alpha_j(k)\in\mathcal{C}_{\alpha_j}$, $j=1,\cdots,n$ 
	is imposed to ensure each agent $j$ moves within its task areas $\mathcal{C}_{\alpha_j}$ while avoiding dangerous positions, such as collisions with industrial plants. In addition, 
	the transition constraint $u(k)=(u_1(k),\cdots,u_n(k))\in\mathcal{C}_u(\alpha(k))$ 
	is imposed to restrict  agents' movements  within the actual velocity limits. Such safety constraints distinguish the model from previous works \cite{hu2022,hu2019}  representing the mobile agent by an unconstrained Markov decision process. 

\subsection{State-dependent fading channel model}
In the considered IIoT system, movements of  agents may lead to shadow fading on the wireless channel 
between sensors and the access point, and finally influence the probability of successful wireless transmission. 
With different time steps, 
during each sampling interval $[t_{\tau k}, t_{\tau(k+1)})$ of the MAS, there are $\tau$  time instants of the wireless transmission as $t_l$, $l={\tau k}, {\tau k+1}, \cdots, {\tau(k+1)-1}$, where the $k$ is determined by the relation $k=\lfloor l/\tau \rfloor$. This means that the successful transmission probabilities for WCS during these $\tau$ consecutive time instants $t_l$ depend on  the state of MAS at time instant $t_{\tau k}$, i.e., $\alpha(k)$.
Then, the WCS and MAS coupling is modeled by the following state-dependent fading channel:
\be\label{eq011}
\mathbb{P}\Big\{\lambda_i(l)=1|\alpha(k)=\alpha,k=\Big\lfloor\frac{l}{\tau}\Big\rfloor\Big\}=\bar{\lambda}_i(\alpha), 
\ee
where  $\lambda_i(l)=1$ indicates the successful transmission  of link $i$ at time instant $t_l$. 

\begin{figure}[!t]
	\centerline{\includegraphics[width=\columnwidth]{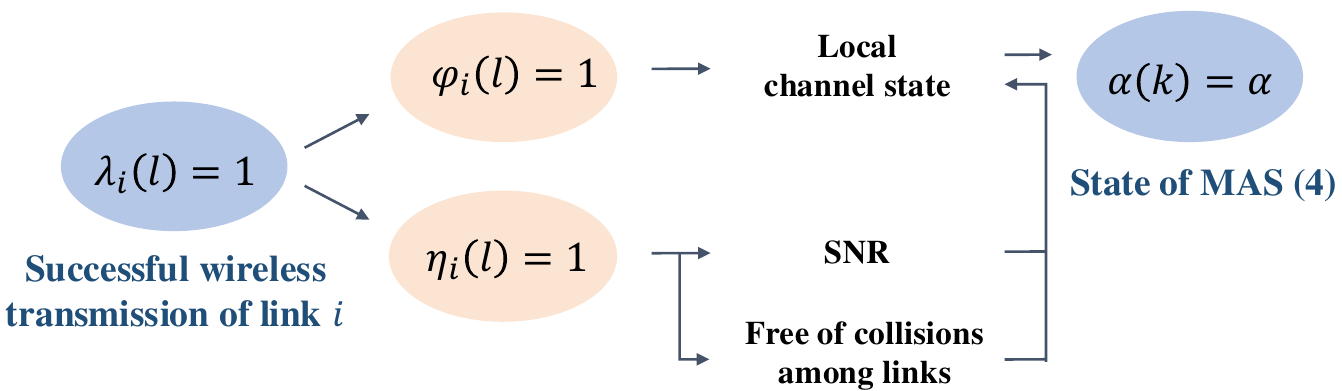}}
	\caption{State-dependent fading channel (\ref{eq011}).}
	\label{fig01-2}
\end{figure}

Let $\eta_i(l)=1$ denote the successful packet decoding of link $i$. As shown in Fig. \ref{fig01-2}, with wireless communication policy (\ref{eq04}), it holds $$\bar{\lambda}_i(\alpha)\!=\!\sum_{c=0}^{r-1}\bar{\gamma}_i(c,\alpha)h_i(c)\mathbb{P}\{\eta_i(l)\!=\!1|\alpha(k)\!=\!\alpha,k=\Big\lfloor\frac{l}{\tau}\Big\rfloor\},$$ 
where $\bar{\gamma}_i(c,\alpha)=\mathbb{P}\{\gamma_i(l)=c|\alpha(k)=\alpha,k=\lfloor l/\tau\rfloor\}$ characterizes the probability distribution for the local channel state,  and can be easily obtained by measuring the local fading amplitude for each MAS state by the statistical methods in  \cite{Agrawal2014,Kashiwagi2010}. Taking into account the interference  among transmission links by adopting the  collision channel model \cite{gatsis2018},  
the last term in $\bar{\lambda}_i(\alpha)$ is equal to $\mathbb{P}\{\hbox{SNR}_i(l)\geq \eta_i^0|\alpha(k)=\alpha,k=\lfloor l/\tau\rfloor,\varphi_i(l)=1,\varphi_j(j)=0,j\neq i\}\prod_{j\neq i}\mathbb{P}\{\varphi_j(l)=0|\alpha(k)=\alpha,k=\lfloor l/\tau\rfloor\}$, where $\hbox{SNR}_i(l)$ 
and  $\eta_i^0$  denote the signal-to-noise ratio 
and the specified decodability threshold, respectively.

Eq. (\ref{eq011}) presents a general state-dependent channel model.  
In this model, the probability of a successful transmission for WCS (\ref{eq01}) depends on the state of MAS (\ref{eq02}). 
In addition, 
as shown in Fig. \ref{fig01-2}, Eq. (\ref{eq011}) takes into account wireless communication policies, the interference among transmission links,  and different time steps for the WCS and MAS, which cannot be captured by  the existing channel models 
 in \cite{hu2022,hu2019}. 
\subsection{Problem formulation}
Considering the state-dependent fading channel model (\ref{eq011}), given an initial state $\alpha_0\in \mathcal{C}_\alpha$, our primary goal is to design an input sequence ${\bf{u}}=\{u(k)\in  \mathcal{C}_u(\alpha(k)): k=0,1,\cdots\}$  for MAS (\ref{eq02}) such that $\alpha(k)\in \mathcal{C}_\alpha$, $k=0,1,\cdots$, and under ${\bf{u}}$ the Lyapunov-like performance 
of the WCS is satisfied, which is formulated as follows.

\begin{defn}\label{def04} The Lyapunov-like performance is said to be guaranteed for WCS (\ref{eq01}) with the state-dependent fading channel (\ref{eq011}), if there exist an input sequence ${\bf{u}}$ and a positive integer $L_0$ such that the inequalities 
	\begin{align}\nonumber
		&\mathbb{E}\Big[V_i(x_i(l+1))|x_i(l),\alpha(k),k=\Big\lfloor\frac{l}{\tau}\Big\rfloor\Big]\\\label{eq014}
		&~~~~~~~~~~~\leq\rho_iV_i(x_i(l))+Tr(Q_i\Xi_i), \forall x_i(l)\in\mathbb{R}^{n_i}
	\end{align}
	hold for any $l\geq L_0$, any $x_i(l)\in\mathbb{R}^{n_i} $ and any $i\in\{1,\cdots,q\}$, where  $V_i(x_i)=x_i^\top Q_ix_i$ is a given quadratic Lyapunov function with $Q_i$ being an $n_i\times n_i$ real positive definite matrix and $\rho_i\in(0,1)$, $i=1,\cdots,q$.
\end{defn}

In (\ref{eq014}), $\rho_i$ represents the  desired decay rate and the trace $Tr(Q_i\Xi_i)$ denotes 
a persistent noise perturbation. Similar kinds of performance requirements as in (\ref{eq014}) have been adopted 
for WCSs \cite{gatsis2015,gatsis2018}. Different from them, Definition \ref{def04} 
explicitly explores the effect of mobile agents' movements on the successful wireless transmission of the WCS. In addition, it allows adopting different time steps to capture different time scales between the  wireless transmission of control systems  and the  movements of mobile agents from cell to cell. 

Our second goal in this paper is to determine optimal input sequence ${\bf{u}}^\ast$ 
such that the average cost function
\begin{align}\nonumber
		&J_{{\bf{u}}}=\lim\limits_{K\rightarrow\infty}\frac{1}{K}\Big\{\sum\limits_{l=0}^{K-1}\sum\limits_{i=1}^{q}\mu_i\mathbb{P}\Big\{\varphi_i(l)=1|\alpha(k),k=\Big\lfloor\frac{l}{\tau}\Big\rfloor\Big\}\\\label{eq012-1}
		&~~~~~~~~~~~~~~~~~~~~~~~+\lambda\sum\limits_{k=0}^{\lfloor (K-1)/\tau\rfloor}g(\alpha(k),u(k))\Big\}
\end{align}
reaches optimality for the initial state $\alpha_0$, where $\lambda$ is a positive constant, $\mu_i$ is the transmit power for sensor $i$, $\sum_{i=1}^{q}\mu_i\mathbb{P}\{\varphi_i(l)=1|\alpha(k),k=\lfloor l/\tau\rfloor\}$ quantifies the power consumption of the WCS, $g: \mathcal{D}_\kappa\times\cdots\times\mathcal{D}_\kappa\rightarrow \mathbb{R}_+$ is the stage cost function of the MAS, and $\mathbb{R}_+$ is the set of positive real numbers, 
 i.e., to solve the infinite-horizon optimal control problem
\begin{align}\nonumber
	P_1:&\min_{{\bf{u}}} J_{{\bf{u}}}\\\nonumber
	&~~\hbox{s.t.}~(\ref{eq01}), (\ref{eq02}),(\ref{eq014}),\\\nonumber
    &~~~~~~~~\alpha(0)=\alpha_0,\\\label{eq013-3}
	&~~~~~~~~\alpha(k)\in \mathcal{C}_\alpha,u(k)\in  \mathcal{C}_u(\alpha(k)), k=0,1,\cdots.
\end{align}



\section{Optimal wireless control as optimal constrained set stabilization}
In the optimal control problem $P_1$, 
there are heterogeneous system dynamics constraints consisting of continuous component (\ref{eq01}) and logical component (\ref{eq02}) with different time steps. In addition, control performance (\ref{eq014}) of the continuous system is constrained by the logical one. The disparity in analyzing these distinct systems makes the resolution of the problem difficult.
To address these challenges, we  convert WCS's performance requirements (\ref{eq014}) together with  safety  constraints (\ref{eq013-3}) 
into the constrained  set stabilization of MAS (\ref{eq02}) using the state-dependent fading channel (\ref{eq011}). 
In this way, problem $P_1$ is converted to the  optimal constrained set stabilization problem of the MAS. 

For the convenience of statement, we call ${\bf{u}}=\{u(k)\in  \mathcal{C}_u(\alpha(k)): k=0,1,\cdots\}$ an  admissible input sequence in the following. 

\begin{defn}\label{def020}  Given $\mathcal{M}\subseteq\mathcal{C}_\alpha$, MAS (\ref{eq02}) is constrained $\mathcal{M}$-stabilizable from initial state $\alpha_0$, if  there exist an  admissible input sequence ${\bf{u}}$ 
	and a positive integer $K_0$ such that $\alpha(k)\in\mathcal{C}_\alpha$ holds  for any $k<K_0$, while $\alpha(k)\in\mathcal{M}$ holds for any $k\geq K_0$.  
\end{defn}

\begin{defn}\label{def03} A subset
	$\mathcal{M}\subseteq \mathcal{C}_\alpha$ is called a constrained control invariant subset of MAS (\ref{eq02}), if for any $\alpha_0\in \mathcal{M}$,
	there exists an admissible input sequence ${\bf{u}}$ 
	such that $\alpha(k)\in\mathcal{M}$ holds for any $k=0,1,\cdots$.
\end{defn}

Obviously, the union of any two constrained control invariant subsets of MAS (\ref{eq02}) is another constrained control invariant subset. The union of all constrained control invariant subsets contained in $\mathcal{M}$ is called its largest constrained control invariant
set, denoted by $I(\mathcal{M})$.

\begin{thm}\label{th01} The Lyapunov-like performance  (\ref{eq014}) is guaranteed for WCS (\ref{eq01}) with the state-dependent fading channel (\ref{eq011}) and safety constraints (\ref{eq013-3}), 
	if and only if MAS (\ref{eq02}) is constrained $I(\Omega(s))$-stabilizable from initial state $\alpha_0$, where $\Omega(s)=\{\alpha\in\mathcal{C}_\alpha:\bar{\lambda}_i(\alpha)\geq s_i, i=1,\cdots,q\}$,  $s=[s_1~\cdots~s_q]$ with 
\be\label{eq053}
s_i=\sup\limits_{y\in\mathbb{R}^{n_i},y\neq 0}\frac{y^\top(A_{o,i}^\top Q_iA_{o,i}-\rho_iQ_i)y}{y^\top(A_{o,i}^\top Q_iA_{o,i}-A_{c,i}^\top Q_iA_{c,i})y}.
\ee
\end{thm}

\proof\ For each subsystem $i$, since $\mathbb{E}[\xi_i(l)]=\mathbf{0}$
and $\lambda_i(l)$ is independent of
$x_i(l)$, by (\ref{eq01}), it holds
\begin{align}\nonumber
	&\mathbb{E}[V_i(x_i(l+1))|x_i(l),\alpha(k),k=\lfloor l/\tau\rfloor]\\\nonumber
	=&\mathbb{P}\{\lambda_i(l)=1|\alpha(k),k=\lfloor l/\tau\rfloor\}x_i^\top(l)A_{c,i}^\top Q_i A_{c,i}x_i(l)\\\nonumber
	&+\mathbb{P}\{\lambda_i(l)=0|\alpha(k),k=\lfloor l/\tau\rfloor\}x_i^\top(l)A_{o,i}^\top Q_i A_{o,i}x_i(l)\\\nonumber
	&+T_r(Q_i\Xi_i).
\end{align}
Then, condition (\ref{eq014}) subject to multiloop WCS dynamics (\ref{eq01}) and safety constraints holds, if and only if there exist an admissible input sequence  and a positive integer $L_0$ such that 
$\alpha(k)\in \mathcal{C}_\alpha$, and for all $x_i(l)\neq \mathbf{0}$, it holds
\begin{align}\nonumber
	&\mathbb{P}\{\lambda_i(l)=1|\alpha(k),k=\lfloor l/\tau\rfloor\}\\\label{eq014-1}
	\geq&\frac{x_i^\top(l)(A_{o,i}^\top Q_i A_{o,i}-\rho_iQ_i)x_i(l)}{x_i^\top(l)(A_{o,i}^\top Q_iA_{o,i}-A_{c,i}^\top Q_iA_{c,i})x_i(l)},
\end{align}
which are equivalent to the condition that there exist an admissible input sequence ${\bf{u}}$ and a positive integer $K_0$, such that $\alpha(k)\in \mathcal{C}_\alpha$, and for all $k\geq K_0$, it holds $\bar{\lambda}_i(\alpha(k))\geq s_i$, $s_i$ is given in (\ref{eq053}), that is, MAS (\ref{eq02}) is constrained $\Omega(s)$-stabilizable from initial state $\alpha_0$.

Next, we prove that MAS (\ref{eq02}) is constrained $\Omega(s)$-stabilizable from initial state $\alpha_0$ if and only if it is constrained $I(\Omega(s))$-stabilizable from initial state $\alpha_0$. We just need to prove the necessity. This is proved by a contradiction argument.  
Since MAS (\ref{eq02}) is constrained $\Omega(s)$-stabilizable from $\alpha_0$, 
we can obtain a feasible admissible input sequence ${\bf{u}}'$ and a positive integer $K_0$, 
such that for any integer $k<K_0$, it holds $\alpha(k;\alpha_0,{\bf{u}}')\in\mathcal{C}_\alpha$; for any integer $k\geq K_0$, it holds $\alpha(k;\alpha_0,{\bf{u}}')\in\Omega(s)$.

Assuming that MAS (\ref{eq02}) is not constrained $I(\Omega(s))$-stabilizable from $\alpha_0$, then under input sequence ${\bf{u}}'$, there exists an integer $k_0\geq K_0$ satisfying
$\alpha(k_0)\notin I(\Omega(s))$. Since $I(\Omega(s))$ is a largest constrained control invariant subset, for any 
$\alpha'\in\Omega(s)$ and any admissible input sequence
${\bf{u}}$, if $\alpha'\notin I(\Omega(s))$, there exists an integer
$k(\alpha',{\bf{u}})$ such that
$\alpha(k(\alpha',{\bf{u}}))\notin\Omega(s)$ from $\alpha'$.  Then, 
under input sequence ${\bf{u}}'$, it holds $\alpha(k_0+k(\alpha(k_0),{\bf{u}}''))\notin\Omega(s)$
from $\alpha_0$, $u''(k)=u'(k+k_0)$, which is a contradiction. Thus, MAS (\ref{eq02}) is constrained $I(\Omega(s))$-stabilizable from  $\alpha_0$.
$\Box$

 By (\ref{eq014-1}),  $s_i$ is the lower bound of the successful transmission probability  for system $i$ ensuring the desired Lyapunov-like performance in Definition \ref{def04}. The value of $s_i$ 
 	can be obtained by the following linear program: 
 	\begin{align}\nonumber
 		&\min_{\theta_i\geq0} ~\theta_i\\\nonumber
 		&~~\hbox{s.t.}~(A_{c,i}^\top Q_iA_{c,i}\!-\! A_{o,i}^\top Q_iA_{o,i})\theta_i\!+\! A_{o,i}^\top Q_iA_{o,i}\!-\! \rho_iQ_i\preceq0,
 	\end{align}
 	which can be solved using the 
 	toolbox such as YALMIP in MATLAB. In addition,  $\Omega(s)$ consists of all constrained states of the MAS satisfying the performance requirements constraint  (\ref{eq014}) of WCS (\ref{eq01}). 

For the optimal control problem $P_1$, all constraints can be guaranteed if MAS (\ref{eq02}) is constrained $I(\Omega(s))$-stabilizable. Since $\mathbb{P}\{\varphi_i(l)=1|\alpha(k)=\alpha,k=\lfloor l/\tau\rfloor\}=\sum_{c=0}^{r-1}\bar{\gamma}_i(c,\alpha)h_i(c)$, ${\bf{u}}^\ast$ is a solution to $P_1$, if and only if it is a solution  to the following 
infinite-horizon optimal constrained set stabilization problem of MAS (\ref{eq02}):
\begin{align}\nonumber
	P_2:&\min_{{\bf{u}}} \bar{J}_{{\bf{u}}}=\lim_{K\rightarrow\infty}\frac{1}{K}\sum_{k=0}^{K-1}\bar{g}(\alpha(k),u(k))\\\nonumber
	&~~\hbox{s.t.}~ (\ref{eq02}), u(k)\in  \mathcal{C}_u(\alpha(k)),  k=0,1,\cdots,\\\nonumber
	&~~~~~~~\,\exists\ K_0>0, \alpha(k)\in I(\Omega(s)),\forall k\geq K_0,\\\nonumber
	&~~~~~~~\, \alpha(k)\in\mathcal{C}_\alpha, \forall k<K_0,
\end{align}
where $\bar{g}(\alpha(k),u(k))$ is the joint stage cost defined as 
$$\bar{g}(\alpha(k),\!u(k))\!=\!\tau\!\sum_{i=1}^q\!\mu_i\!\sum_{c=0}^{r-1}\!\bar{\gamma}_i(c, \alpha(k))h_i(c)\!+\!\lambda g(\alpha(k),\!u(k)).$$
Let the optimal objective values of problems $P_1$ and $P_2$ be $J^
\ast=J_{{\bf{u}}^\ast}$ and $\bar{J}^
\ast=\bar{J}_{{\bf{u}}^\ast}$, respectively. It is easy to obtain that $J^
\ast=\bar{J}^
\ast/\tau$.

\section{Infinite-horizon optimal controller design}

Note that both the state space and control space of the MAS dynamics (\ref{eq02}) are  finite. This property allows us to adopt the
	graphical method to address the optimization problem $P_2$. 
	By reformulating the logical-state MAS in a tractable form via the semi-tensor
	product approach, the constrained optimal state transition graph is firstly  constructed to encode the constraints and the objective function of  $P_2$. 
	Based on this, the feasibility of  problem $P_2$ is deduced to the existence of  cycles in the graph, and  the design of optimal input sequences is modeled as a minimum-mean cycle problem. An algorithm is then proposed to  construct optimal controllers.
\subsection{Feasibility analysis via a graphical method}

Due to the highly nonlinear finite-field model, 
	it is difficult to apply classical nonlinear system theory to investigate (\ref{eq02}). Recently, the semi-tensor product approach
	has emerged 
	as an elegant tool to deal with such issues \cite{chengqi2011,cheng2011}. The basic idea is to embed the nonlinear finite-filed dynamics in (\ref{eq02}) into a higher dimensional space where its evolution is linear. 
	Specially, representing $i-1\in\mathcal{D}_\kappa$ as the vector $\delta_\kappa^i$, $i=1,\cdots,\kappa$, then $\alpha_i(k),u_i(k)\in\mathcal{D}_\kappa$ have their corresponding
	vector forms (still using the same symbols) $\alpha_i(k),u_i(k)\in Col(I_\kappa)$,
	respectively.  
	According to similar calculation to \cite{wangs2022}, we convert MAS (\ref{eq02}) into  the following bilinear equivalent algebraic form: \be\label{eq015}
	\alpha(k+1)=Fu(k)\alpha(k),
	\ee
	where  
	$\alpha(k)=\ltimes_{i=1}^n \alpha_i(k)\subseteq Col(I_N)$, $u(k)=\ltimes_{i=1}^n u_i(k)\subseteq Col(I_N)$,   $N:=\kappa^n$, and $F$ is a logical matrix \cite{chengqi2011,cheng2011} characterizing all state-transition information of the MAS. Then,  the set of admissible  inputs steering MAS (\ref{eq015}) from $\delta_N^a$ to $\delta_N^b$ in one step can be obtained as
\be\label{eq001-1}
U_{a,b}=\{\delta_N^l\in \mathcal{C}_u(\delta_N^a): [F\delta_N^l]_{b,a}=1\}.
\ee In addition, the set of  optimal inputs enabling the transition with the lowest cost is \be\label{eq001}
\bar{U}_{a,b}={\arg\min}_{\delta_N^l\in U_{a,b}}\bar{g}(\delta_N^a,\delta_N^l).
\ee 




\begin{figure}[!t]
	\centerline{\includegraphics[width=3in]{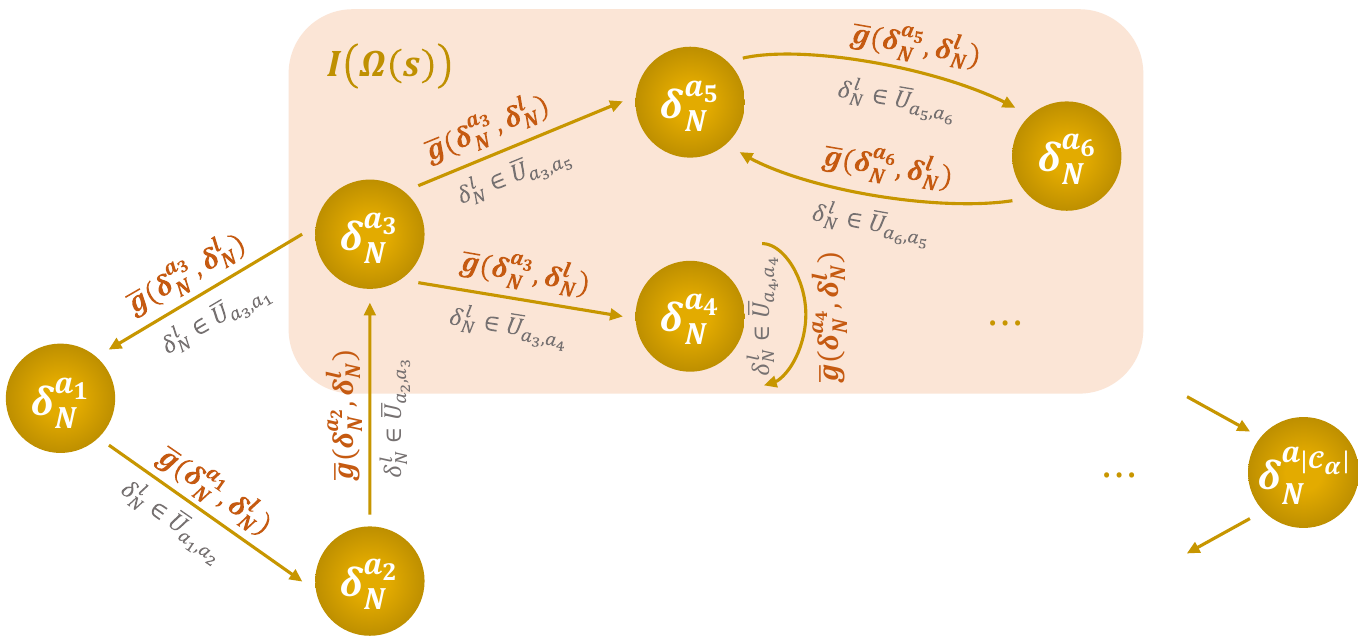}}
	\caption{Constrained optimal state transition graph $G=(\mathcal{V},\mathcal{E},w)$.}
	\label{graph}
\end{figure}
To determine the feasibility of optimization problem $P_2$, we construct the constrained optimal state transition graph $G=(\mathcal{V},\mathcal{E},w)$ to encode  constraints and the objective function.  
Regard each state $\delta_N^a\in\mathcal{C}_\alpha$ 
as a vertex on a directed graph,
and encode each constrained state transition from $\delta_N^a$ to $\delta_N^b$ into edge $(\delta_N^a,\delta_N^b)$. 
Additionally,  assign the weight  $w(\delta_N^a,\delta_N^b)=\bar{g}(\delta_N^a,\delta_N^l)$ to the edge $(\delta_N^a,\delta_N^b)\in\mathcal{E}$ to encode the lowest joint cost under the optimal input  $\delta_N^l\in \bar{U}_{a,b}$. 
See Fig. \ref{graph} for an illustration.

 For an input sequence ${\bf{u}}$, we say that the state trajectory steered by it converges to a cycle $\mathcal{C}$, if the path $\mathcal{T}=\{\alpha(k): k=0,1,\cdots\}$ in $G$  keeps repeating $\mathcal{C}$ after a finite time. 
Define the weight and average weight of a path  $\mathcal{T}=\{\delta_N^{a_0},\cdots,\delta_N^{a_l}\}\subseteq\mathcal{V}$, $l\geq 0$ as $w(\mathcal{T})=\sum_{i=0}^{l-1}w(\delta_N^{a_i},\delta_N^{a_{i+1}})$ and $\bar{w}(\mathcal{T})=w(\mathcal{T})/l$, respectively.
Then, we have the following result. 

\begin{thm}\label{lem001} MAS (\ref{eq015}) is constrained $I(\Omega(s))$-stabilizable from initial state $\alpha_0$, if and only if $I(\Omega(s))\cap \mathcal{R}(\alpha_0)\neq\emptyset$, where $\mathcal{R}(\alpha_0)$ denotes the set of constrained states reachable from $\alpha_0$. 
	In addition, if  MAS (\ref{eq015}) is constrained $I(\Omega(s))$-stabilizable from  $\alpha_0$, then 
	\begin{enumerate}
		\item[\rm (i)] it holds $C(G[\Phi])\neq\emptyset$, where $\Phi=I(\Omega(s))\cap \mathcal{R}(\alpha_0)$, $G[\Phi]=(\Phi,\mathcal{E}',w)$ is the induced subgraph of $G$ induced by  $\Phi$ with $\mathcal{E}'=\{(\delta_N^a,\delta_N^b)\in \mathcal{E}: \delta_N^a,\delta_N^b\in \Phi\}$, and $C(\cdot)$ represents  the set of all cycles in the graph; 
		\item[\rm (ii)] solutions to the infinite-horizon optimal control problem $P_2$ exist.
	\end{enumerate}
\end{thm}

\proof\ We just need to prove the sufficiency. If $I(\Omega(s))\cap \mathcal{R}(\alpha_0)\neq\emptyset$, then there exist $\alpha'\in I(\Omega(s))$, $K>0$ and an input sequence ${\bf{u}}_1=\{u(k)\in  \mathcal{C}_u(\alpha(k)): k=0,\cdots,K-1\}$  such that from initial state  $\alpha_0$,  $\alpha(K)=\alpha'$ and  $\alpha(k)\in\mathcal{C}_\alpha$ holds for any $k<K$. In addition, since $I(\Omega(s))$ is a constrained control invariant subset, there exists an admissible input sequence ${\bf{u}}_2$ such that from initial state  $\alpha'$,  $\alpha(k)\in I(\Omega(s))$, $k=0,1,\cdots$. Thus, by Definition \ref{def020}, ${\bf{u}}=\{{\bf{u}}_1,{\bf{u}}_2\}$ is a feasible input sequence such that MAS (\ref{eq015}) is constrained $I(\Omega(s))$-stabilizable from $\alpha_0$.

Since $I(\Omega(s))$ is a constrained control invariant subset, for any initial state $\alpha''\in\Phi$, there exists an admissible  input sequence ${\bf{u}}_1$ 
such that $\alpha(k)\in I(\Omega(s))$ holds for any $k=0,1,\cdots,|I(\Omega(s))|$. Then, there exist $0\leq k_1$, $k_2\leq |I(\Omega(s))|$, $k_1\neq k_2$ satisfying  $\alpha(k_1)=\alpha(k_2)$. Since $\alpha''\in\mathcal{R}(\alpha_0)$, we obtain a cycle in $G[\Phi]$ as $\{\alpha(k):k=k_1,\cdots,k_2\}$, that is, $C(G[\Phi])\neq\emptyset$. Thus, (i) is proved.

According to the proof of (i), if MAS (\ref{eq015}) is constrained $I(\Omega(s))$-stabilizable from initial state $\alpha_0$, we can obtain a feasible admissible input sequence ${\bf{u}}_2$ and a cycle $\mathcal{C}\in C(G[\Phi])$ such that the state trajectory from $\alpha_0$ steered by ${\bf{u}}_2$ converges to  $\mathcal{C}$.  Consider the corresponding path $\mathcal{T}=\{\alpha(k):k=0,\cdots,K\}$ in $G$. Since we focus on the infinite-horizon optimal control problem, consider the case
where $K$ is big enough such that $\mathcal{T}$ contains at least one cycle $\mathcal{C}$. Then, path $\mathcal{T}$ can always be decomposed to a finite subpath before entering $\mathcal{C}$, $r=\frac{1}{|\mathcal{C}|-1}(K-|\mathcal{T}_1|-|\mathcal{T}_2|+2)$ cycles $\mathcal{C}$ and a  subpath  of $\mathcal{C}$, denoted by $\mathcal{T}_2$, which can be an empty set. Then, we have $w(\mathcal{T})=w(\mathcal{T}_1)+r(|\mathcal{C}|-1)\bar{w}(\mathcal{C})+w(\mathcal{T}_2)$.
Since $w(\mathcal{T}_1)$ and $w(\mathcal{T}_2)$ are bounded, it holds
\begin{align}\nonumber
	\lim\limits_{K\rightarrow\infty}\frac{w(\mathcal{T})}{K}=&\lim\limits_{K\rightarrow\infty}\frac{K-|\mathcal{T}_1|-|\mathcal{T}_2|+2}{K(|\mathcal{C}|-1)}(|\mathcal{C}|-1)\bar{w}(\mathcal{C})\\\label{eq005}
	=&\bar{w}(\mathcal{C}).
\end{align}
On the other hand, by (\ref{eq012-1}), we have
	$J_{{\bf{u}}_2}=\lim_{K\rightarrow\infty}w(\mathcal{T})$ $/K$, 
which together with (\ref{eq005}) shows that under the feasible admissible input sequence ${\bf{u}}_2$, the limit of the cost function exist. Thus, solutions to the infinite-horizon optimal control problem $P_2$ exist.
$\Box$

For the feasibility of optimal constrained set stabilization problem $P_2$, the remaining problem is how to construct $\Phi$.
Algorithm \ref{al02} is presented for the calculation of $I(\Omega(s))$. The main idea  is to recursively eliminate states in $\Omega(s)$ with no successor state inside $\Omega(s)$ (Line $2$) until no states to be removed (Line 3). The reason is that  the trajectories cannot be kept inside  $\Omega(s)$ starting from such states. It is clear that Algorithm \ref{al02} terminates in at most $\Omega(s)$ recursions. 
\begin{algorithm}
	\caption{\textbf{: Calculation of largest constrained control invariant subset $I(\Omega(s))$}}
	\label{al02}
	\begin{algorithmic}[1]
		\Require $\Omega(s)$, $F$
		\Ensure $I(\Omega(s))$
		\State $\Lambda_0\leftarrow \Omega(s)$, $i\leftarrow 1$
		\State Calculate $\Lambda_i=\{\alpha\in\Lambda_{i-1}:\mathcal{R}_1(\alpha)\cap\Lambda_{i-1}\neq\emptyset\}$
		\If{$\Lambda_i==\Lambda_{i-1}$}
		$I(\Omega(s))\leftarrow\Lambda_i$ and Break 
		\Else\ 
		$i\leftarrow i+1$ and go to Line 2
		\EndIf
	\end{algorithmic}
\end{algorithm}

Denote the set of constrained states reachable from $\alpha_0$ in at least $K$ time steps by $\mathcal{R}_K(\alpha_0)=\{\alpha\in\mathcal{C}_\alpha:  \alpha(K)=\alpha, \exists \hbox{an admissible input sequence}\ {\bf{u}}$, \hbox{and}\  $\alpha(k)\in\mathcal{C}_\alpha, \alpha\notin\mathcal{R}_{k}(\alpha_0), \forall k< K\}$, $K\geq 1$, and $\mathcal{R}_0(\alpha_0)=\{\alpha_0\}$. Then, one can obtain 
$$\mathcal{R}_1(\delta_N^a)=\{Col_a(F\delta_N^l),  \delta_N^l\in \mathcal{C}_u(\delta_N^a)\}\cap\mathcal{C}_\alpha.$$
Since there are at most $|\mathcal{C}_\alpha|-1$ reachable states from $\alpha_0$, one can obtain $\mathcal{R}(\alpha_0)=\cup_{K=1}^{l}\mathcal{R}_K(\alpha_0)$, where $l$ is a positive integer satisfying $l\leq |\mathcal{C}_\alpha|-1$. Based on this idea, 
Algorithm \ref{al03} is presented to calculate $\mathcal{R}(\alpha_0)$.
\begin{algorithm}[htp]
	\caption{\textbf{: Calculation of constrained reachable set $\mathcal{R}(\alpha_0)$}}
	\label{al03}
	\begin{algorithmic}[1]
		\Require $\mathcal{R}_0(\alpha_0)=\{\alpha_0\}$
		\Ensure $\mathcal{R}(\alpha_0)$
		\State  $\mathcal{R}(\alpha_0)\leftarrow \emptyset$, $K\leftarrow 1$
		\State Calculate 
		$\mathcal{R}_K(\alpha_0)=\{\alpha\in\cup_{\alpha\in\mathcal{R}_{K-1}(\alpha_0)}\mathcal{R}_1(\alpha):\alpha\notin\mathcal{R}(\alpha_0)\}$
		\State $\mathcal{R}(\alpha_0)\leftarrow \mathcal{R}(\alpha_0)\cup \mathcal{R}_K(\alpha_0)$
		\If{$\mathcal{R}_K(\alpha_0)==\emptyset$} Break 
		\Else\ 
		$K\leftarrow K+1$ and go to Line 2
		\EndIf
	\end{algorithmic}
\end{algorithm}

\subsection{Optimal controller design}
To improve the computational efficiency of problem solving, motivated by \cite{gao2021-1,Zhao2011,zhu2018}, we convert  the infinite-horizon optimal constrained set stabilization problem $P_2$ into a minimum-mean cycle problem  in $G$. Then,  by analyzing the properties of the graph, an efficient algorithm is proposed to design optimal input sequences. 

\begin{lem}\label{lem002} If  MAS (\ref{eq015}) is constrained $I(\Omega(s))$-stabilizable from initial state $\alpha_0$, then there exists a simple cycle $\mathcal{C}^\ast$ which is a minimum-mean cycle in $C(G[\Phi])$, that is, $\mathcal{C}^\ast=\min_{\mathcal{C}\in C(G[\Phi])}\bar{w}(\mathcal{C})$.
\end{lem}
\proof\ According to Theorem \ref{lem001}, we have $C(G[\Phi])\neq\emptyset$.  Assume  $\mathcal{C}^\ast$ is a minimum-mean cycle in $G[\Phi]$. If $\mathcal{C}^\ast$ does not have any other repeated vertices except the first and last vertices, then the conclusion is proved. Otherwise, decompose $\mathcal{C}^\ast$ into simple cycles as $\mathcal{C}_1^\ast, \mathcal{C}_2^\ast$, $\cdots,\mathcal{C}_l^\ast$.
Since 
\be\label{eq002}
\bar{w}(\mathcal{C}^\ast)=\frac{\sum_{i=1}^l\bar{w}(\mathcal{C}_{i}^\ast)(|\mathcal{C}_{i}^\ast|-1)}{\sum_{i=1}^l(|\mathcal{C}_{i}^\ast|-1)},
\ee
we have $\bar{w}(\mathcal{C}^\ast)\geq\min_{1\leq i\leq l}\bar{w}(\mathcal{C}_{i}^\ast)$, which together with the fact that $\mathcal{C}^\ast$ is a minimum-mean cycle in $G[\Phi]$ shows that $\bar{w}(\mathcal{C}^\ast)=\min_{1\leq i\leq l}\bar{w}(\mathcal{C}_{i}^\ast)$. 
Thus, one can obtain $\bar{w}(\mathcal{C}_i^\ast)=\bar{w}(\mathcal{C}^\ast)$, $i=1,\cdots,l$.
$\Box$

The following  result presents the property of optimal input sequences. 

\begin{thm}\label{thm002} Assume that  MAS (\ref{eq015}) is constrained $I(\Omega(s))$-stabilizable from initial state $\alpha_0$. 
An input sequence ${\bf{u}}^\ast=\{u^\ast(k):k=0,1,\cdots\}$ is optimal, 
if the state trajectory steered by it converges to a simple minimum-mean cycle $\mathcal{C}^\ast$ in $G[\Phi]$, and each input $u^\ast(k)$ belongs to the optimal input set $\bar{U}_{a,b}$ in (\ref{eq001}) after entering $\mathcal{C}^\ast$. In addition, it holds $J^\ast=\bar{w}(\mathcal{C}^\ast)/\tau$.
\end{thm}

\proof\ According to the proof of (ii) in Theorem \ref{lem001} and Lemma \ref{lem002},  it holds $\bar{J}_{{\bf{u}}^\ast}=\bar{w}(\mathcal{C}^\ast)$.  Given any feasible input sequence ${\bf{u}}=\{u(k):k=0,\cdots,K-1\}$ of problem $P_2$, consider the path $\mathcal{T}=\{\alpha(k):k=0,\cdots,K\}$ in $G$, where $K$ is big enough such that $\mathcal{T}$  contains at least one cycle. Assume that $\alpha(k)\in\Phi$ holds for any $k\geq K_0$ and denote the trajectory before and after entering $\Phi$ as $\mathcal{T}_1=\{\alpha(k):k<K_0\}$ and $\mathcal{T}_2=\{\alpha(k):k\geq K_0\}$, respectively. Then,  there exists $l\geq 1$ such that 
$$\frac{w(\mathcal{T})}{K}=\frac{w(\mathcal{T}_1)}{K}+\frac{1}{K}\sum_{i=1}^{l}\bar{w}(\mathcal{C}_i)(|\mathcal{C}_i|-1)+\frac{w(\mathcal{T}_3)}{K},$$
where $\mathcal{T}_1$ is a finite path, $\mathcal{C}_i$, $i=1,\cdots,l$ are simple cycles in $G[\Phi]$, and $\mathcal{T}_3$ is a simple path in $G[\Phi]$ with $\mathcal{E}(\mathcal{T}_2)$ being the union of $\mathcal{E}(\mathcal{C}_1), \cdots, \mathcal{E}(\mathcal{C}_l)$,  $\mathcal{E}(\mathcal{T}_3)$ preserving duplications. Since $\mathcal{C}^\ast$ is a simple minimum-mean cycle, we have 
\begin{align}\nonumber
	\frac{w(\mathcal{T})}{K}&\geq\frac{w(\mathcal{T}_1)}{K}+\frac{1}{K}\sum_{i=1}^{l}\bar{w}(\mathcal{C}^\ast)(|\mathcal{C}_i|-1)+\frac{w(\mathcal{T}_3)}{K},\\\nonumber
	&=\frac{w(\mathcal{T}_1)}{K}+\frac{K-K_0-|\mathcal{T}_3|+2}{K}\bar{w}(\mathcal{C}^\ast)+\frac{w(\mathcal{T}_3)}{K}.
\end{align}
Since $w(\mathcal{T}_1)$ and $w(\mathcal{T}_3)$ are bounded, it holds
\begin{align}\nonumber
	\bar{J}_{{\bf{u}}}=&\frac{1}{K}\sum_{k=0}^{K-1}\bar{g}(\alpha(k),u(k))\\\nonumber
	\geq&\lim\limits_{K\rightarrow\infty}\!\Big(\!\frac{K-K_0-|\mathcal{T}_3|+2}{K}\bar{w}(\mathcal{C}^\ast)+\frac{w(\mathcal{T}_1)}{K}+\frac{w(\mathcal{T}_3)}{K}\!\Big)\\\nonumber
	=&\bar{w}(\mathcal{C}^\ast),
\end{align}
where the first inequality follows from (\ref{eq001}).
Thus, for any feasible input sequence ${\bf{u}}$, it holds $\bar{J}_{{\bf{u}}}\geq \bar{J}_{{\bf{u}}^\ast}$, that is, $\bar{J}^\ast
=\bar{w}(\mathcal{C}^\ast)$. Therefore, we have $J^\ast=\bar{w}(\mathcal{C}^\ast)/\tau$.
$\Box$

Under the control of ${\bf{u}}^\ast$, there exists $K_0>0$ such that $\alpha(K_0)\in\mathcal{C}^\ast\subseteq\Omega(s)$, $k\geq K_0$. Then, according  Theorem \ref{th01}, the Lyapunov-like performance  (\ref{eq014}) is guaranteed for WCS (\ref{eq01}) with the state-dependent fading channel (\ref{eq011}). In addition, according to Theorem \ref{thm002}, if $I(\Omega(s))$ is a singleton, we further have the following result.

\begin{cor}\label{cor} Assume that  MAS (\ref{eq015}) is constrained $I(\Omega(s))$-stabilizable from initial state $\alpha_0$ and $I(\Omega(s))=\{\alpha_e\}$. Any feasible input sequence  ${\bf{u}}^\ast=\{u^\ast(k): k=0,1,\cdots\}$ is optimal, where each input $u^\ast(k)$, $k> K_0$ belongs to the optimal input set $\bar{U}_{a,b}$ in (\ref{eq001}). In addition, $\alpha(k)=\alpha_e$ holds for any $k\geq K_0$, and $J^\ast(\alpha_0)=\bar{w}(\{\alpha_e,\alpha_e\})/\tau$.
\end{cor}

In order to construct optimal input sequences ${\bf{u}}^\ast$, the remaining problem is how to locate minimum-mean cycles in a directed weighted graph.  For  $G[\Phi]$, let $o\in\Phi$ be a source vertex that can reach every vertex  in $\Phi$. Denote by $H(k,\alpha)$ the minimum  weight of any $k$-edge path from $o$ to $\alpha\in\Phi$. If no such path exists, $H(k,\alpha):=\infty$. According to \cite{Chaturvedi2017}, if the average weight of the minimum-mean cycle in $G[\Phi]$ is $\epsilon^\ast=\bar{w}(\mathcal{C}^\ast)$, then it holds 
\be\label{eq003}
\epsilon^\ast=\min\limits_{\alpha\in\Phi}\max\limits_{0\leq k\leq |\Phi|-1}\frac{H({|\Phi|},\alpha)-H(k,\alpha)}{|\Phi|-k}.
\ee
Let $\alpha^\ast$ and $k^\ast$ be an optimal solution to (\ref{eq003}). Then, every cycle on the $|\Phi|$-edge path from $o$ to $\alpha^\ast$ of weight $H({|\Phi|},\alpha^\ast)$ is a minimum-mean cycle.
If the source vertex $o$ does not exist in $G[\Phi]$, then partition $G[\Phi]$ into strongly connected components, and find the minimum-mean cycle  in each strongly connected component \cite{karp1978}. 

\begin{algorithm}
	\caption{\textbf{: Construction of optimal input sequence ${\bf{u}}^\ast$}}
	\label{al04}
	\begin{algorithmic}[1]
		\Require $G[\Phi]$, $\alpha_0=\delta_N^{a_0}$, $\mathcal{R}_i(\alpha_0)$, $i=1,\cdots,r$
		\Ensure ${\bf{u}}^\ast$ 
		\State Compute all strongly connected components of $G[\Phi]$ as $G[\mathcal{V}_i]=(\mathcal{V}_i,\mathcal{E}_i,w)$, $i=1,\cdots,\tau$ 
		\For {$i\leftarrow 1$ to $\tau$}
		\State Select an arbitrary source vertex $o_i$ from $\mathcal{V}_i$
		\State Solve (\ref{eq003}) to obtain $\epsilon_i^\ast$, 
		$\alpha_i^\ast$ via dynamic program- $~~~~~~$ming \cite{karp1978}
		\EndFor
		\State $\iota\leftarrow\arg\min_{1\leq i\leq\tau}\epsilon_i^\ast$
		\State Find a $|\mathcal{V}_\iota|$-edge path $p_\iota^\ast$ from $o_\iota$ to $\alpha_\iota^\ast$ of weight $H(|\mathcal{V}_\iota|,\alpha_\iota^\ast)$
		\State Find a simple cycle $\mathcal{C}^\ast$ on the path $p_\iota^\ast$, $l\leftarrow|\mathcal{C}^\ast|-1$
		\If{$\alpha_0\in\mathcal{C}^\ast$}  
		\State  Rearrange the element of $\mathcal{C}^\ast$ as $\mathcal{C}^\ast=\{\delta_N^{c^\ast[0]}$, $~~~~~~$$\delta_N^{c^\ast[1]}, \cdots,\delta_N^{c^\ast[l]}\}$ such that $c^\ast[0]=c^\ast[l]=a_0$ 
		\State $\varsigma\leftarrow 0$, 
		 go to Line $17$
		\Else\   $\varsigma\leftarrow\arg\min_{i=1}^r\{\mathcal{R}_i(\alpha_0)\cap\mathcal{C}^\ast\neq\emptyset\}$
		\State Create an array $t^\ast$ of size $\varsigma+1$ with $t^\ast[0]\leftarrow a_0$ $~~~~~~$and $t^\ast[\varsigma]\leftarrow a$, $\delta_N^a\in\mathcal{R}_\varsigma(\alpha_0)\cap\mathcal{C}^\ast$
		\For {$k\leftarrow \varsigma-1$ to $1$}
		\State $t^\ast[k]\leftarrow a$, $\delta_N^a\in\mathcal{R}_k(\alpha_0)$, $\delta_N^{t^\ast[k+1]}\in\mathcal{R}_1(\delta_N^a)$
		\EndFor
		\State The trajectory before enter $\mathcal{C}^\ast$ is $\mathcal{T}^\ast=\{\delta_N^{t^\ast[0]}$, $~~~~~~$$\delta_N^{t^\ast[1]}$, $\cdots,\delta_N^{t^\ast[\varsigma-1]}\}$
		\State Rearrange the element of $\mathcal{C}^\ast$ as $\mathcal{C}^\ast=\{\delta_N^{c^\ast[0]}$, $~~~~~~$$ \delta_N^{c^\ast[1]}$, $\cdots,\delta_N^{c^\ast[l]}\}$ such that $c^\ast[0]=c^\ast[l]=t^\ast[\varsigma]$
		\EndIf
		\State An optimal input sequence is ${\bf{u}}^\ast=\{u^\ast(k): k=0,1,\cdots\}$ with 
		$$u^\ast(k)\!\in\!\left\{
		\begin{array}{lll}
			U_{t^\ast[k],t^\ast[k+1]}, ~~~~~~~~~~~~~~~~~k=0,\cdots,\varsigma-1;\\
			\bar{U}_{c^\ast[k-(\varsigma+jl)],c^\ast[k-(\varsigma+jl)+1]},~~~~~~k=\varsigma+jl,\\
			~~~~~~~~\,~\cdots,\varsigma+(j+1)l-1, ~j=0,1,\cdots,
		\end{array}
		\right.$$
		where $U_{a,b}$, $\bar{U}_{a,b}$ are respectively defined in (\ref{eq001-1}), (\ref{eq001})  
		\State Recover the logical form of 
		 the input sequence ${\bf{u}}^\ast$
	\end{algorithmic}
\end{algorithm}

Algorithm \ref{al04} is proposed for the construction of optimal input sequences. The procedure computes the  strongly connected components of $G[\Phi]$ by Tarjan's algorithm (Line 1). Within each component, solve Eq. (\ref{eq003})   by dynamic programming \cite{karp1978} (Lines 2-4). And find a simple  minimum-mean cycle $\mathcal{C}^\ast$ in $G[\Phi]$ by keeping track of the vertices in a path with a backpointer (Lines 5-7).  Then, according to the constrained reachable sets and the breath-first search algorithm, find the trajectory from $\alpha_0$ before entering $\mathcal{C}^\ast$  as $\mathcal{T}^\ast$ (Lines 11-15). In addition,  rearrange the elements of $\mathcal{C}^\ast$ to connect $\mathcal{T}^\ast$ and $\mathcal{C}^\ast$ (Line 16). Finally, design optimal input sequence to drive $\alpha_0$ to approach and repeat $\mathcal{C}^\ast$ (Line 17), and recover  the logical form of the obtained optimal input sequence via the procedure in  \cite{chengqi2011,cheng2011} (Line 18). 

\section{An illustrative example}
Two autonomous assembly arms and two AGVs coordinate  to jointly execute the assembly process in a circular workspace with a radius of $3$ meters. AGVs  transport the raw materials, subassemblies and products, while the autonomous assembly arms perform the  manufacturing operations on the line.

\begin{table*}[htb]\centering
	\caption{Fading channel parameters.}
	\label{table1}
	\begin{tabular}{ccccccccccccccccc}
		\toprule[0.8pt]
		$(i,c)$&\multicolumn{2}{c}{\begin{footnotesize}$\small{(1,0)}$\end{footnotesize}} &\multicolumn{2}{c}{\begin{footnotesize}$(1,1)$ \end{footnotesize}} & \multicolumn{2}{c}{\begin{footnotesize}$(1,2)$\end{footnotesize}} &\multicolumn{2}{c}{\begin{footnotesize}$(1,3)$\end{footnotesize}} &\multicolumn{2}{c}{\begin{footnotesize}$(2,0)$\end{footnotesize}} &\multicolumn{2}{c}{\begin{footnotesize}$(2,1)$ \end{footnotesize}} & \multicolumn{2}{c}{\begin{footnotesize}$(2,2)$\end{footnotesize}} &\multicolumn{2}{c}{\begin{footnotesize}$(2,3)$\end{footnotesize}}  \\ \hline
		\begin{footnotesize}$\delta_9^1$\end{footnotesize} & \begin{footnotesize}0.18\end{footnotesize} & \begin{footnotesize}\textbf{0.0}\end{footnotesize}         & \begin{footnotesize}0.18\end{footnotesize}        & \begin{footnotesize}\textbf{0.2}\end{footnotesize}         & \begin{footnotesize}0.18\end{footnotesize}        & \begin{footnotesize}\textbf{0.1}\end{footnotesize}         &
		\begin{footnotesize} 0.18\end{footnotesize}        & \begin{footnotesize}\textbf{0.7}\end{footnotesize}        & \begin{footnotesize}0.44\end{footnotesize}        & \begin{footnotesize}\textbf{0.2}\end{footnotesize}         & \begin{footnotesize}0.44\end{footnotesize}        & \begin{footnotesize}\textbf{0.1}\end{footnotesize}         & \begin{footnotesize}0.44\end{footnotesize}        & \begin{footnotesize}\textbf{0.5}\end{footnotesize}         &
		\begin{footnotesize} 0.44\end{footnotesize}        & \begin{footnotesize}\textbf{0.2}\end{footnotesize}  \\
		\begin{footnotesize}$\delta_9^2,\delta_9^4$\end{footnotesize}  & \begin{footnotesize}0.66\end{footnotesize}        & \begin{footnotesize}\textbf{0.1}\end{footnotesize}         & \begin{footnotesize}0.66\end{footnotesize}        & \begin{footnotesize}\textbf{0.1}\end{footnotesize}         & \begin{footnotesize}0.66\end{footnotesize}        & \begin{footnotesize}\textbf{0.3}\end{footnotesize}         & \begin{footnotesize}0.66\end{footnotesize}        & \begin{footnotesize}\textbf{0.5}\end{footnotesize}        & \begin{footnotesize}0.37\end{footnotesize}        & \begin{footnotesize}\textbf{0.2}\end{footnotesize}         & \begin{footnotesize}0.37\end{footnotesize}        & \begin{footnotesize}\textbf{0.1}\end{footnotesize}         & \begin{footnotesize}0.37\end{footnotesize}        & \begin{footnotesize}\textbf{0.1}\end{footnotesize}         &
		\begin{footnotesize} 0.37\end{footnotesize}        & \begin{footnotesize}\textbf{0.6}\end{footnotesize}  \\ 
		\begin{footnotesize}$\delta_9^3,\delta_9^7$\end{footnotesize}  & \begin{footnotesize}0.18\end{footnotesize}        & \begin{footnotesize}\textbf{0.2}\end{footnotesize}         & \begin{footnotesize}0.18\end{footnotesize}        & \begin{footnotesize}\textbf{0.1}\end{footnotesize}         & \begin{footnotesize}0.18\end{footnotesize}        & \begin{footnotesize}\textbf{0.2}\end{footnotesize}         & \begin{footnotesize}0.18\end{footnotesize}        & \begin{footnotesize}\textbf{0.5}\end{footnotesize}        & \begin{footnotesize}0.31\end{footnotesize}        & \begin{footnotesize}\textbf{0.2}\end{footnotesize}         & \begin{footnotesize}0.31\end{footnotesize}        & \begin{footnotesize}\textbf{0.1}\end{footnotesize}         & \begin{footnotesize}0.31\end{footnotesize}        & \begin{footnotesize}\textbf{0.5}\end{footnotesize}         &
		\begin{footnotesize} 0.31\end{footnotesize}        & \begin{footnotesize}\textbf{0.2}\end{footnotesize}  \\ 
		\begin{footnotesize}$\delta_9^5$\end{footnotesize}  & \begin{footnotesize}0.64\end{footnotesize}        & \begin{footnotesize}\textbf{0.3}\end{footnotesize}         & \begin{footnotesize}0.64\end{footnotesize}        & \begin{footnotesize}\textbf{0.1}\end{footnotesize}         & \begin{footnotesize}0.64\end{footnotesize}        & \begin{footnotesize}\textbf{0.2}\end{footnotesize}         & \begin{footnotesize}0.64\end{footnotesize}        & \begin{footnotesize}\textbf{0.4}\end{footnotesize}        & \begin{footnotesize}0.25\end{footnotesize}        & \begin{footnotesize}\textbf{0.0}\end{footnotesize}         & \begin{footnotesize}0.25\end{footnotesize}        & \begin{footnotesize}\textbf{0.2}\end{footnotesize}         & \begin{footnotesize}0.25\end{footnotesize}        & \begin{footnotesize}\textbf{0.1}\end{footnotesize}         &
		\begin{footnotesize} 0.25\end{footnotesize}        & \begin{footnotesize}\textbf{0.6}\end{footnotesize}  \\ 
		\begin{footnotesize}$\delta_9^6,\delta_9^8$\end{footnotesize}  & \begin{footnotesize}0.54\end{footnotesize}        & \begin{footnotesize}\textbf{0.3}\end{footnotesize}         & \begin{footnotesize}0.54\end{footnotesize}        & \begin{footnotesize}\textbf{0.1}\end{footnotesize}         & \begin{footnotesize}0.54\end{footnotesize}        & \begin{footnotesize}\textbf{0.2}\end{footnotesize}         & \begin{footnotesize}0.54\end{footnotesize}        & \begin{footnotesize}\textbf{0.4}\end{footnotesize}        & \begin{footnotesize}0.30\end{footnotesize}        & \begin{footnotesize}\textbf{0.2}\end{footnotesize}         & \begin{footnotesize}0.30\end{footnotesize}        & \begin{footnotesize}\textbf{0.1}\end{footnotesize}         & \begin{footnotesize}0.30\end{footnotesize}        & \begin{footnotesize}\textbf{0.1}\end{footnotesize}         &
		\begin{footnotesize} 0.30\end{footnotesize}        & \begin{footnotesize}\textbf{0.6}\end{footnotesize}  \\ 
		\begin{footnotesize}$\delta_9^9$\end{footnotesize}  & \begin{footnotesize}0.18\end{footnotesize}        & \begin{footnotesize}\textbf{0.3}\end{footnotesize}         & \begin{footnotesize}0.18\end{footnotesize}        & \begin{footnotesize}\textbf{0.1}\end{footnotesize}         & \begin{footnotesize}0.18\end{footnotesize}        & \begin{footnotesize}\textbf{0.2}\end{footnotesize}         & \begin{footnotesize}0.18\end{footnotesize}        & \begin{footnotesize}\textbf{0.4}\end{footnotesize}        & \begin{footnotesize}0.25\end{footnotesize}        & \begin{footnotesize}\textbf{0.2}\end{footnotesize}         & \begin{footnotesize}0.25\end{footnotesize}        & \begin{footnotesize}\textbf{0.1}\end{footnotesize}         & \begin{footnotesize}0.25\end{footnotesize}        & \begin{footnotesize}\textbf{0.5}\end{footnotesize}         &
		\begin{footnotesize} 0.25\end{footnotesize}        & \begin{footnotesize}\textbf{0.2}\end{footnotesize}  \\ 
		\bottomrule [0.8pt]
	\end{tabular}\\
	\footnotetext[1] 01. The numbers in bold denote $\bar{\gamma}_i(c,\alpha)$, while the others denote $\bar{\eta}_i(\alpha)$.
\end{table*}

The workspace is partitioned into three two dimensional cells. The two autonomous assembly arms are located in Cell $0$ and Cell $2$, respectively.  Given that the sensors operate at a sampling rate of 20 Hz, they track the predefined desired trajectories by exchanging information between autonomous assembly arms and remote controllers via a shared wireless medium. The dynamics of the autonomous assembly arms 
are in the form of	(\ref{eq01}) with states $x_1(k)=[x_{11}(k)~x_{12}(k)]^\top\in\mathbb{R}^{2}$ and $x_2(k)\in\mathbb{R}$
denoting the differences between current and desired positions,
where
$$A_{c,1}=\left[
\begin{array}{cc}
	-0.1 & -0.1 \\
	~~0.1 & ~~0.2 \\
\end{array}
\right],~
A_{o,1}=\left[
\begin{array}{cc}
	~-1 & -0.4 \\
	-0.5 & ~~0.3 \\
\end{array}
\right]$$
and $A_{c,2}=0.2$, $A_{o,2}=1$. They are perturbed by zero-mean unite-variance Gaussian noises. Let $\rho_1=0.95$, $\rho_2=0.9$,  $Q_1$ solve
$A_{c,1}^\top Q_1A_{c,1}-Q_1+I=0$ and $Q_2=1$. Then, by (\ref{eq053}), the lower bounds of the probability of successful transmission that ensure the desired Lyapunov decay rates are $s_1\approx 0.29$
and $s_2\approx 0.10$.

The dynamics of  two AGVs in the continuous space are $z_j(k+1)=z_j(k)+2\nu_j(k)$, $j=1,2$, i.e., the time steps are $2$ s. 
Give the waypoints  centered in $(0,2)$, $(-2,0)$, $(2,0)$ for AGV $1$, and centered in $(0,1)$, $(-\sqrt{3}/2,-0.5)$, $(\sqrt{3}/2,-0.5)$ for AGV $2$. The  maximum speeds of AGVs are $1.5$ meters per second. 
The finite-field model of the MAS 
is in the form of (\ref{eq02}) with $a_{1,1}=a_{2,1}=a_{2,2}=1$ and $a_{1,2}=2$.
AGV $1$  and $2$ respectively start from Cells $1$ and $0$. 
The task areas for AGVs $1, 2$ are respectively given as $\{\hbox{Cell}\ 0$, $\hbox{Cell}\ 1\}$ and  $\{\hbox{Cell}\ 0$, $\hbox{Cell}\ 1, \hbox{Cell}\ 2\}$. In addition, due to the velocity limits, AGV $1$ cannot move between the waypoints  in $\hbox{Cell}\ 1$ and $\hbox{Cell}\ 2$ within one time step.
  Then, in vector form, the initial state of the MAS is $\alpha_0=\delta_{9}^{4}$, and safety constraints are $\mathcal{C}_\alpha=\{\delta_9^1,\cdots,\delta_9^6\}$, $\mathcal{C}_u(\alpha(k))=\{\delta_9^4,\delta_9^5,\delta_9^7,\delta_9^8\}$, $\forall \alpha(k)$.


For the wireless channel that the autonomous assembly arms are controlled over, let $s=4$, the pre-designed wireless communication policy be $h_i(0)=h_i(1)=h_i(2)=1$, $h_i(3)=0$, $i=1, 2$ and the transmit power be $\mu_1=0.25$, $\mu_2=0.5$. 
The dependence of local channel states on the MAS's state, i.e., $\bar{\gamma}_i(c,\alpha)$, and the dependence of successfully decoded probability on the MAS's state, i.e., $\bar{\eta}_{i}(\alpha)$, are shown in Table \ref{table1}.  Then, the state-dependent fading channel (\ref{eq011}) can be obtained as $\Lambda_1=[0.05~0.33~0.09~0.33~0.38~0.32~0.09~0.32~0.11]$ and
$\Lambda_2=[0.35~0.15~0.25~0.15~0.10~0.12~0.25~0.12~0.20]$, where $Col_a(\Lambda_i)=\bar{\lambda}_i(\delta_9^a)$, $a=1,\cdots,9, i=1,2$. 

By the state-dependent channel,  the set of all constrained states of the MAS ensuring the desired Lyapunov decay rates is
$\Omega(s)=\{\delta_{9}^2,\delta_{9}^4,\delta_{9}^5,\delta_{9}^6\}$. The largest constrained control invariant subset of $\Omega(s)$ is $I(\Omega(s))=\Omega(s)$. Calculate the constrained reachable set $\mathcal{R}(\alpha_0)=\mathcal{C}_\alpha$. Since $I(\Omega(s))\cap \mathcal{R}(\alpha_0)\neq\emptyset$, by Theorem \ref{th01} and Theorem \ref{lem001}, the Lyapunov-like performance is guaranteed for the WCS with the state-dependent fading channel and safety constraints.


\begin{figure}[!t]
	\centerline{\includegraphics[width=4cm]{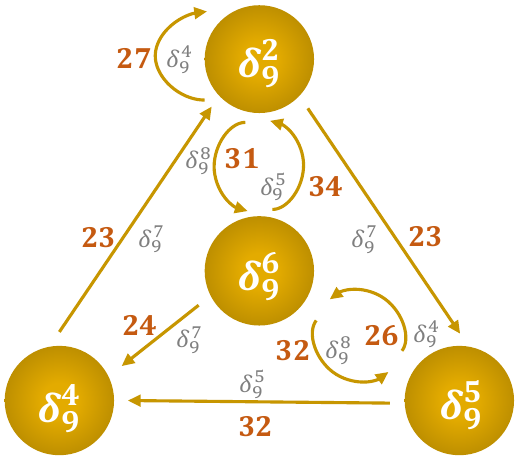}}
	\caption{Weighted state transition
		graph $G[\Phi]$.}
	\label{fig03}
\end{figure}

\begin{figure}[!t]
	\centering
	\subfigure[]{
		\includegraphics[width=0.22\textwidth]{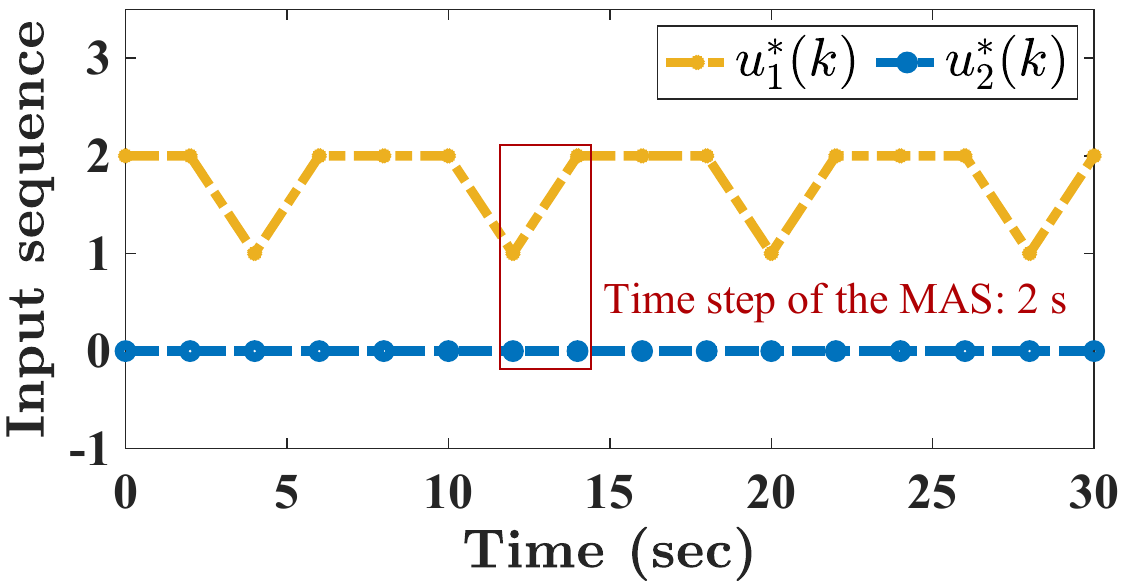}
	}
	\label{fig09-1}
	\subfigure[]{
		\includegraphics[width=0.22\textwidth]{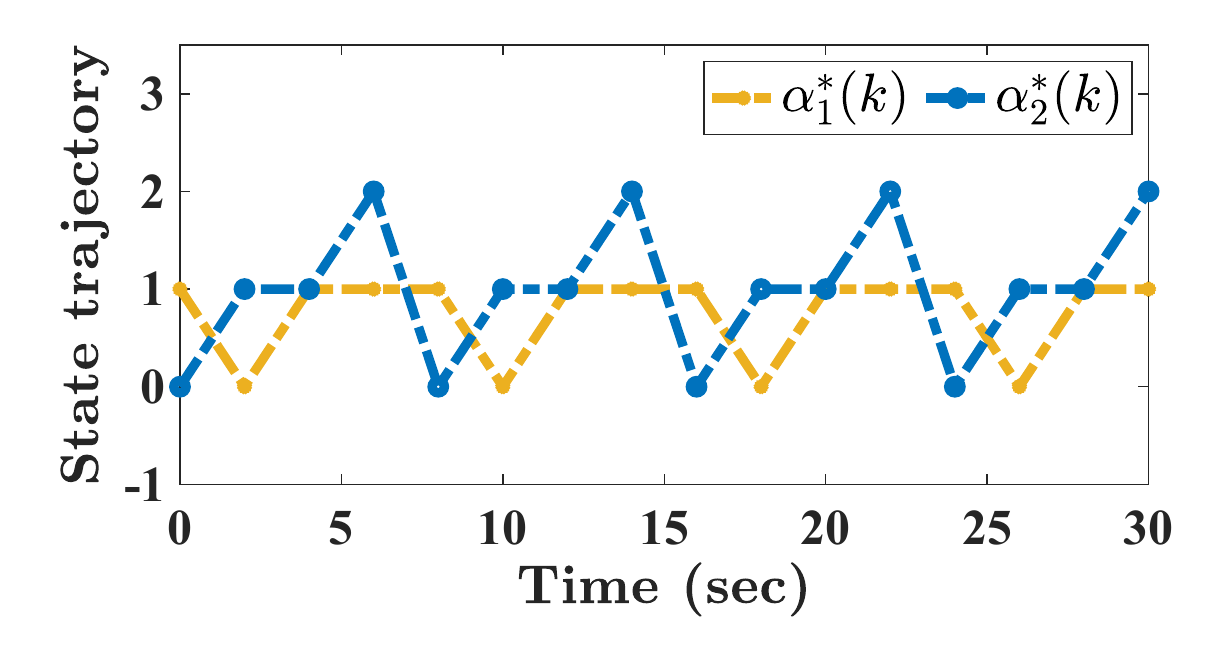}
	}
	\label{fig09-2}
	\subfigure[]{
		\includegraphics[width=0.22\textwidth]{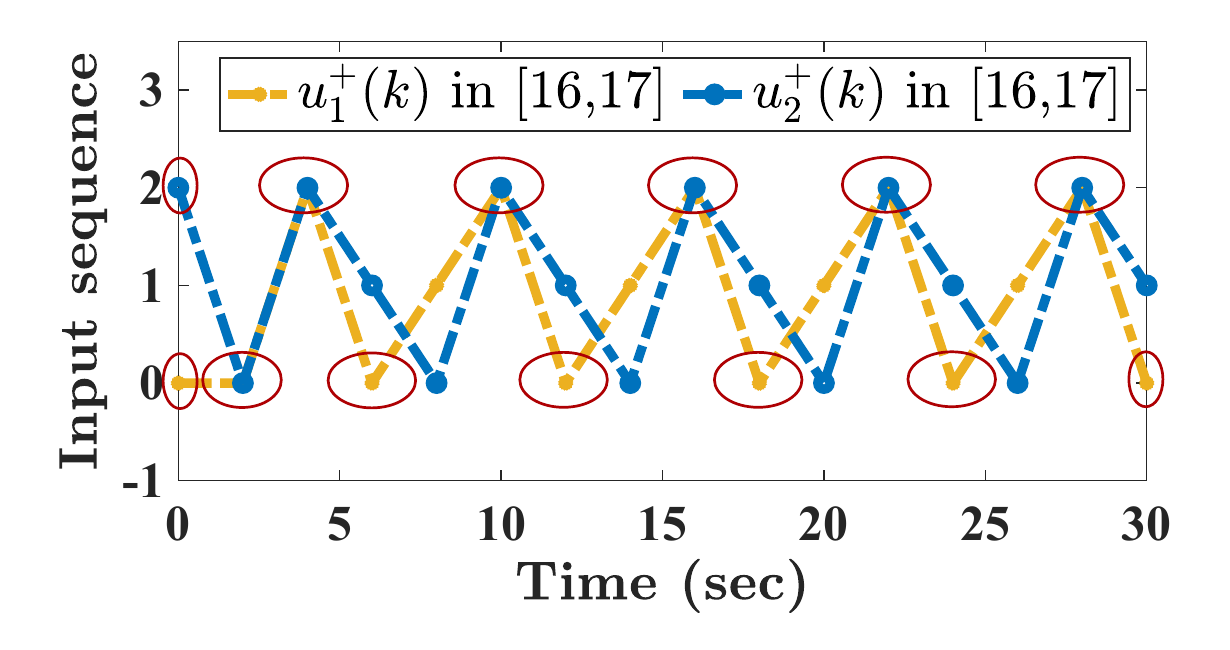}
	}
	\label{fig09-3}
	\subfigure[]{
		\includegraphics[width=0.22\textwidth]{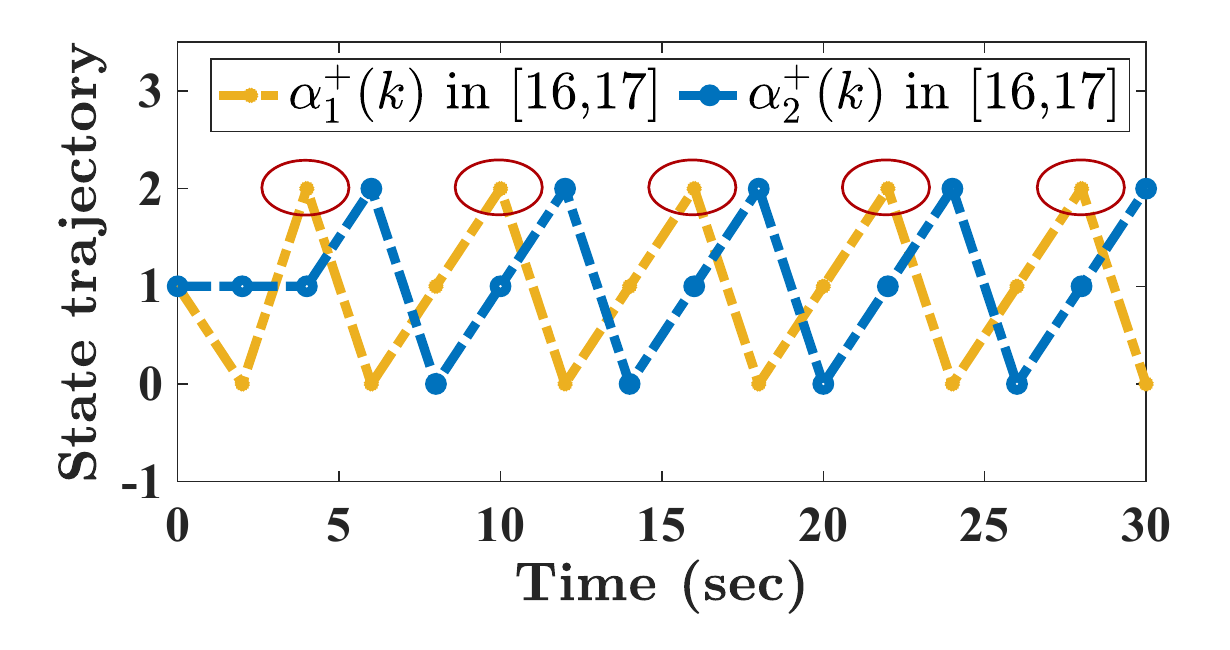}
	}
	\label{fig09-4}
	\caption{Input sequences and steered state trajectories for AGVs in logical forms in the time interval $0\!-\!30$ s, where (a), (b) are obtained by Algorithm \ref{al04},  (c), (d) are obtained by the method proposed in \cite{hu2022,hu2019}, states and control inputs violating safety constraints are depicted with red circles.}
	\label{optimal}
\end{figure}

For the infinite-horizon average cost,  let $\lambda=1$ and
$C=[C_1~\cdots~C_9]$ with $C_l=[8~10~12~14~20~14~10~18~16]$, $l=1,\cdots,9$, where $Col_a(C_l)=g(\delta_9^a,\delta_9^l)$, $a,l=1,\cdots,9$ are stage costs. The weighted state transition
graph $G[\Phi]$ with $\Phi=I(\Omega(s))$ is shown in Fig. \ref{fig03}. It is easy to see that there exist eight simple constrained cycles in $G[\Phi]$ as
\begin{align}\nonumber
	&\mathcal{C}_1=\{\delta_9^2,\delta_9^2\}, \mathcal{C}_2=\{\delta_9^2,\delta_9^6,\delta_9^2\}, \mathcal{C}_3=\{\delta_9^5,\delta_9^6,\delta_9^5\},\\\nonumber
	&\mathcal{C}_4=\{\delta_9^2,\delta_9^6,\delta_9^4,\delta_9^2\}, \mathcal{C}_5=\{\delta_9^2,\delta_9^5,\delta_9^6,\delta_9^2\}, \\\nonumber
	&\mathcal{C}_6=\{\delta_9^2,\delta_9^5,\delta_9^4,\delta_9^2\},\mathcal{C}_7=\{\delta_9^2,\delta_9^6,\delta_9^5,\delta_9^4,\delta_9^2\},\\\nonumber &\mathcal{C}_8=\{\delta_9^2,\delta_9^5,\delta_9^6,\delta_9^4,\delta_9^2\}.
\end{align}
 Since $\bar{w}(\mathcal{C}_8)=24=\min_{i\in\{1,\cdots,8\}}\bar{w}(\mathcal{C}_i)$,
according to Algorithm \ref{al04}, an optimal input sequence can be designed as
${\bf{u}}^\ast=\{u^\ast(k): k=0,1,\cdots\}$ with $$u^\ast(k)=\left\{
\begin{array}{lll}
	\delta_9^4, & k=2,6,10, \cdots;\\
	\delta_9^7, & \hbox{otherwise}.
\end{array}
\right.$$
The optimal objective value of problem $P_2$ is $\bar{J}_{{\bf{u}}^\ast}=\bar{w}(\mathcal{C}_8)=24$. Since $\tau=40$, according to Theorem \ref{thm002}, it holds $J^\ast=0.6$. Under the control of ${\bf{u}}^\ast$, the  state trajectory of the MAS is $\mathcal{T}=\{\delta_{9}^{4},\delta_{9}^{2},\delta_{9}^{5},\delta_{9}^{6},\delta_{9}^{4},\delta_{9}^{2},\delta_{9}^{5},\delta_{9}^{6}$, $\cdots\}$. We respectively plot in Fig. \ref{optimal} (a)-(b) the optimal input sequences and state trajectories for AGVs in logical forms in the time interval $0\!-\!30$ s. Especially, Fig. \ref{optimal} (a) shows that the update of the state for AGV $1$  is driven  by its in-neighbor's dynamics, i.e.,  AGV $2$, and its control input, while that for AGV $2$  is only driven  by its in-neighbor's dynamics, i.e.,  AGV $1$. 
	With the same  fading channel parameters, wireless communication and transmission power policies, based on the method proposed in \cite{hu2022,hu2019}, a feasible state-feedback control policy is $u^+(k)=\pi(\alpha(k))$ with 
	$\pi(0,0)=(1,1)$, $\pi(0,1)=(0,0)$, $\pi(0,2)=(0,1)$, $\pi(1,1)=(0,2)$, $\pi(2,1)=(2,2)$, and $\pi(i,j)=(1,0)$, $(i,j)\in\{(1,0),(1,2),(2,0),(2,2)\}$. Under this control policy, the input sequences and steered state trajectories are respectively shown in Fig. \ref{optimal} (c)-(d). 
As shown in Fig. \ref{optimal}, Algorithm \ref{al04} takes into account safety constraints of the MAS while \cite{hu2022,hu2019} does not.

\begin{figure}[!t]
	\centerline{\includegraphics[width=6.8cm]{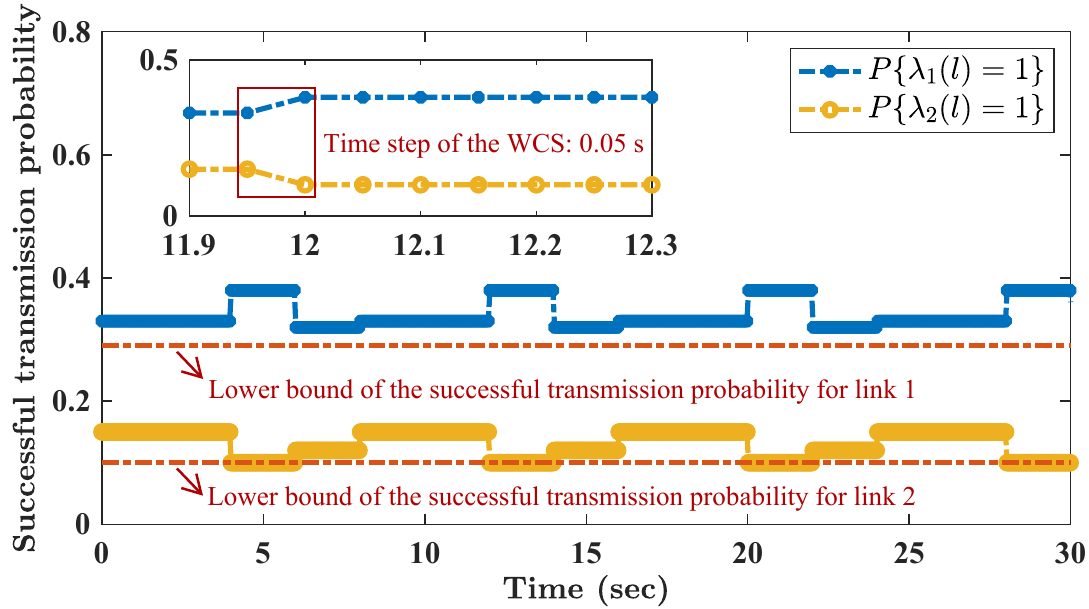}}
	\caption{Successful transmission probabilities for the two links in the time interval $0\!-\!30$ s.}
	\label{successful transmission probability}
\end{figure}

\begin{figure}[!t]
	\centerline{\includegraphics[width=6.8cm]{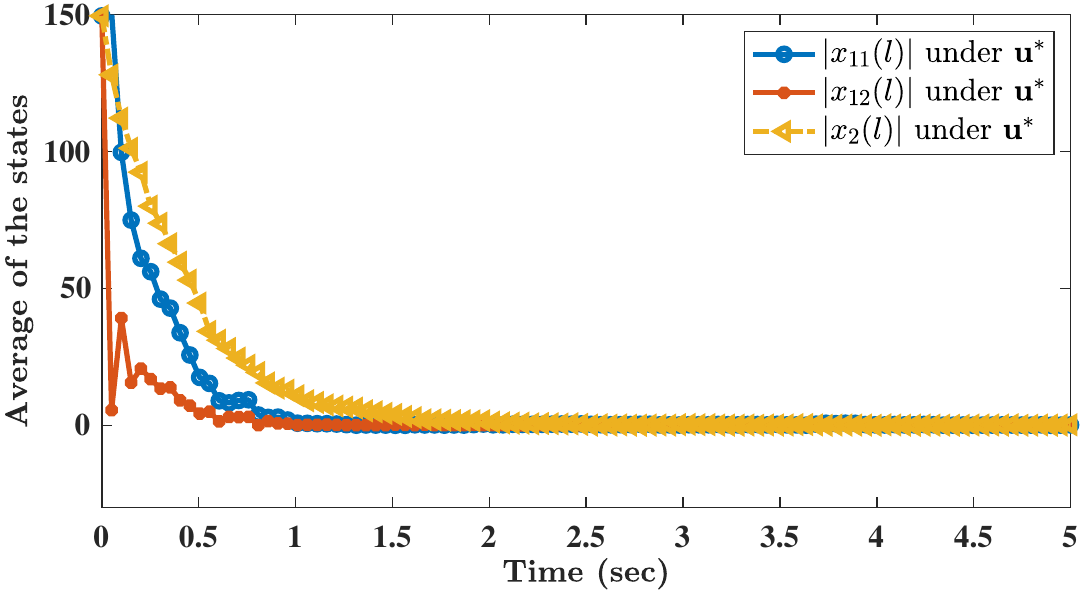}}
	\caption{The evolution of the empirical averages of the states for WCS in the time interval $0\!-\!5$ s under state feedback law ${\bf{u}}^\ast$.}
	\label{fig05}
\end{figure}

With the state trajectory $\mathcal{T}$ of the MAS,  we respectively plot in Fig. \ref{successful transmission probability} the  successful transmission probabilities for the two links in the time interval $0\!-\!30$ s under the effect of state-dependent shadow fading. 
Since $\mathbb{P}\{\lambda_i(l)$ $=1|\alpha(k),k=\lfloor l/{\tau}\rfloor\}\geq s_i$, $l\geq 0$, $i=1,2$, the designed input sequence ${\bf{u}}^\ast$ ensures the desired Lyapunov performance requirements. Following the Monte Carlo method, we plot in Fig. \ref{fig05}
the evolution of the empirical averages of the states for WCS in the time interval $0\!-\!5$ s under the obtained optimal input sequence ${\bf{u}}^\ast$. 

\begin{figure}[!t]
	\centerline{\includegraphics[width=6.8cm]{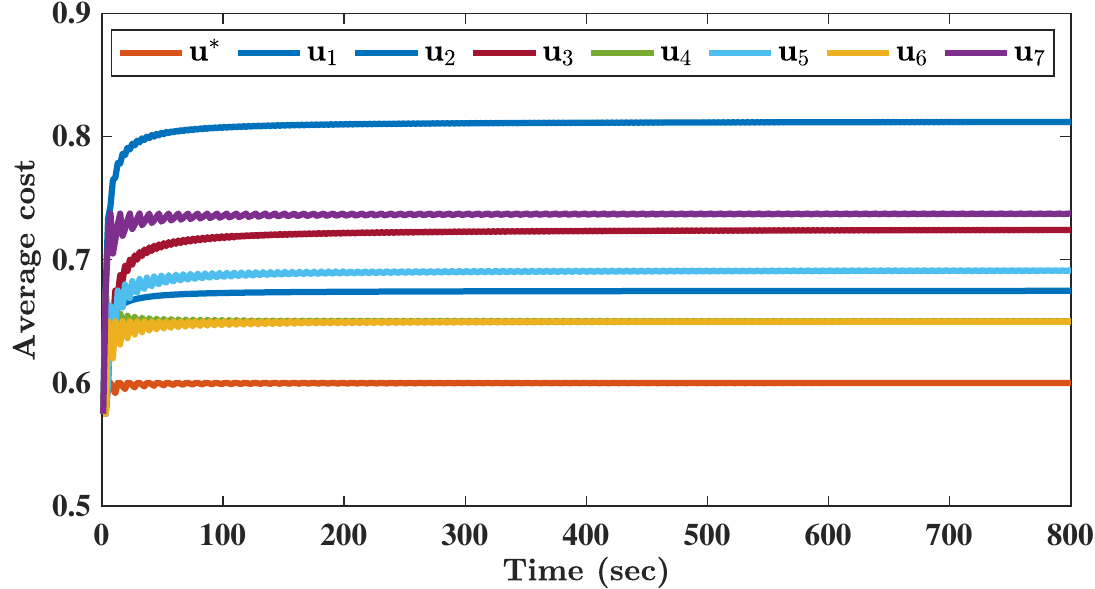}}
	\caption{Average cost in the time interval $0\!-\!800$ s under input sequences ${\bf{u}}^\ast$ and ${\bf{u}}_i$, $i=1$, $\cdots,7$, respectively.}
	\label{fig06}
\end{figure}

\begin{figure}[!t]
	\centerline{\includegraphics[width=5cm]{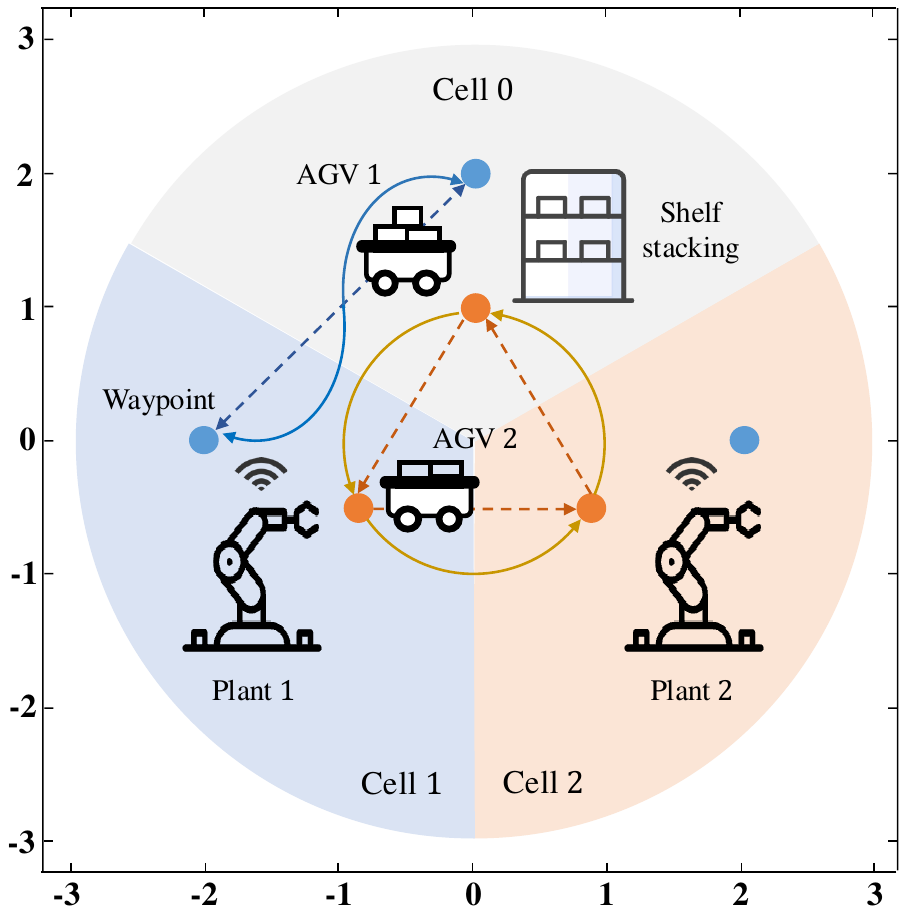}}
	\caption{
			Straight  and curved  AGVs paths in  the continuous space following optimal state trajectories at the high level, depicted with directed dashed and solid  lines, respectively.}
	\label{example}
\end{figure}

We respectively plot in Fig. \ref{fig06}
the average cost  of problem $P_1$ in the time interval $0\!-\!800$ s associated with optimal input sequence ${\bf{u}}^\ast$, and random feasible input sequences ${\bf{u}}_i$, $i=1,\cdots,7$ under which the state trajectory enters $\mathcal{C}_i$.
Simulation results show that optimal input sequence can ensure the Lyapunov-like performance of the WCS with lower average cost. Moreover, the infinite-horizon average cost is determined by the average cost of the cycle that the state converges to. Thus, the obtained input sequence ${\bf{u}}^\ast$ solves the optimal control problem $P_1$. 
	By respectively designing velocities $\nu_1$ and $\nu_2$  for AGVs $1$ and $2$,  Fig.  \ref{example}  plots two commonly used types of AGVs paths  in  the continuous space  
	following the optimal state trajectories at the high level, i.e., straight  and curved paths. Other advanced control strategies can also be employed for control design.


\section{Conclusion}
In this paper, we have focused on the infinite-horizon optimal control of MAS to ensure the performance of multiloop WCS, while minimizing an average joint cost for the IIoT system in the presence of WCS and MAS coupling.  We have characterized  both the interference among transmission links and the coupling between WCS and MAS by a state-dependent fading channel. 
Then, the optimal control problem is formulated as the optimal constrained set stabilization for the MAS to address hybrid state spaces and different time steps. Finally, by constructing a constrained optimal state transition graph, an effective algorithm is proposed to construct optimal input sequences for the
	MAS  based on minimum-mean cycles for
	weighted graphs. 

For the future work, it is an important and interesting topic to consider the co-design problem of transmission scheduling and mobile agents' motion control to further improve energy efficiency. 
Future work includes also the incorporation of  safety constraints for WCSs and high-level intelligent mission of MASs.


\begin{thebibliography}{1}

\bibitem{Abl:56}
B.C. Able.
\newblock Nucleic acid content of microscope.
\newblock {\em Nature}, 135:7--9, 1956.

\bibitem{AbTaRu:54}
B.C. Able, R.A. Tagg, and M.~Rush.
\newblock Enzyme-catalyzed cellular transanimations.
\newblock In A.F. Round, editor, {\em Advances in Enzymology}, volume~2, pages
  125--247. Academic Press, New York, 3rd edition, 1954.

\bibitem{Keo:58}
R.~Keohane.
\newblock {\em Power and Interdependence: World Politics in Transitions}.
\newblock Little, Brown \& Co., Boston, 1958.

\bibitem{Pow:85}
T.~Powers.
\newblock Is there a way out?
\newblock {\em Harpers}, pages 35--47, June 1985.

\bibitem{Heritage:92}
A.~H. Soukhanov, editor.
\newblock {\em {The American Heritage. Dictionary of the American Language}}.
\newblock Houghton Mifflin Company, 1992.

\end{thebibliography}


\begin{thebibliography}{99}     

\bibitem{Agrawal2014}  Agrawal, P.,  Ahl\'{e}n, A.,  Olofsson, T., \&  Gidlund, M. (2014). Long term channel characterization for energy efficient transmission in industrial environments, {\it IEEE Trans. Commun., 62(8)}, 3004--3014.

\bibitem{Ahlen2019}  Ahl\'{e}n, A.,  Akerberg, J.,  Eriksson, M.,  Isaksson, A. J.,  Iwaki, T.,  Johansson, K. H.,  Knorn, S.,  Lindh, T., \&  Sandberg, H. (2019). Toward wireless control in industrial process automation: A case study at a paper mill, {\it IEEE Control Syst. Mag., 39(5)}, 36--57.
     
\bibitem{Baumann2021} Baumann, D., Mager, F., Wetzker, U., Thiele, L., Zimmerling, M., \& Trimpe, S.  (2021). Wireless control for smart manufacturing: Recent approaches and open challenges, {\it Proc. IEEE, 109(4)}, 441--467.

\bibitem{Chaturvedi2017}  Chaturvedi, M., \&  McConnell, R. M. (2017). A note on finding minimum mean cycle, {\it Inf. Process. Lett., 127},  21--22.

\bibitem{chengqi2011} Cheng, D.,   \& Qi, H. (2011). {\it Semi-tensor Product of Matrices: Theory and Application, 2nd Edition}, in Chinese. Beijing, China: Science Press.

\bibitem{cheng2011}  Cheng, D.,  Qi, H., \& Zhao, Y. (2012). {\it An Introduction to Semi-tensor Product of Matrices and Its Applications}.  Singapore: World Scientific.

\bibitem{donkers2011}  Donkers, M.,  Heemels, W., van de Wouw, N., \&  Hetel, L. (2011). Stability analysis of networked control systems using a switched linear systems approach, {\it IEEE Trans. Autom. Control, 56(9)}, 2101--2115.

\bibitem{fadlullah2011} Fadlullah, Z. M.,  Fouda, M. M.,  Kato, N.,  Takeuchi, A.,  Iwasaki, N., \&  Nozaki, Y. (2011). Toward intelligent machine-to-machine communications in smart grid, {\it IEEE Commun. Mag., 49(4)}, 60--65.

\bibitem{Fornasini2014}  Fornasini, E., \&  Valcher, M. E. (2014). Optimal control of Boolean control networks,  {\it IEEE Trans. Autom. Control, 59(5)},  1258--1270.

\bibitem{gao2021-1}  Gao, S.,  Sun, C.,  Xiang, C.,  Qin, K., \&  Lee, T. H. (2022). Infinite-horizon optimal control of switched Boolean control networks with average cost: An efficient graph-theoretical approach, {\it IEEE Trans. Cybern., 52(4)}, 2314--2328.


\bibitem{gatsis2015} Gatsis, K.,  Pajic, M.,  Ribeiro, A., \&  Pappas, G. (2015). Opportunistic control over shared wireless channels, {\it IEEE Trans. Autom. Control, 60(12)}, 3140--3155.

\bibitem{Gatsis2014}  Gatsis, K.,  Ribeiro, A., \&  Pappas, G.  (2014). Optimal power management in wireless control systems, {\it IEEE Trans. Autom. Control, 59(6)},  1495--1510.

\bibitem{gatsis2018}  Gatsis, K.,  Ribeiro, A., \&  Pappas, G. (2018). Random access design for wireless control systems, {\it Automatica,  91},  1--9.

\bibitem{Hristu2001}  Hristu-Varsakelis, D. (2001). Feedback control systems as users of a shared network: Communication sequences that guarantee stability, In  {\it Proc. 40th IEEE Conference on Decision and Control}, Orlando, USA (pp. 3631--3636).


\bibitem{hu2014}  Hu, B., \&  Lemmon, M. D. (2014). Event triggering in vehicular networked systems with limited bandwidth and deep fading, In {\it Proc. 53rd IEEE Conference on Decision and Control}, California, USA (pp. 3542--3547).



\bibitem{hu2022}  Hu, B., \&  Tamba, T. A. (2022). Optimal transmission power and controller design for networked control systems under state-dependent Markovian channels, {\it IEEE Trans. Autom. Control, 67(10)}, 5669--5676.

\bibitem{hu2019}  Hu, B.,  Wang, Y.,  Orlik, P. V.,  Koike-Akino, T., \&  Guo, J. (2019). Co-design of safe and efficient networked control systems in factory automation with state-dependent wireless fading channels, {\it Automatica, 105}, 334--346.

\bibitem{hu2011}  Hu, Y., \&  Ribeiro, A., (2011). Adaptive distributed algorithms for optimal random access channels, {\it IEEE Trans. Wireless Commun., 10(8)}, 2703--2715.

\bibitem{karp1978} Karp, R. (1978). A characterization of the minimum cycle mean in a digraph, {\it Discr. Math., 23(3)}, 309--311.

\bibitem{Kashiwagi2010}  Kashiwagi, I.,  Taga, T., \&  Imai, T. (2010). Time-varying path-shadowing model for indoor populated environments, {\it IEEE Trans. Veh. Technol., 59(1)}, 16--28.

\bibitem{le2011}  Le Ny, J.,  Feron, E., \&  Pappas, G. J. (2011). Resource constrained LQR control under fast sampling, In {\it Proc. 14th International Conference on Hybrid Systems: Computation and Control}, Chicago, USA (pp. 271--280).

\bibitem{Leong2016}  Leong, A. S.,  Quevedo, D. E.,  Ahl\'{e}n, A., \&  Johansson, K. H. (2016). On network topology reconfiguration for remote state estimation, {\it IEEE Trans. Autom. Control, 61(12)}, 3842--3856.

\bibitem{Liy2019}  Li, Y.,  Li, H.,   \&  Ding, X. (2020). Set stability of switched delayed logical networks with application to finite-field consensus, {\it Automatica, 113}, 108768.


\bibitem{Lidl1996}  Lidl, R., \&  Niederreiter, H. (1996). {\it Finite Fields}. New York, NY, USA: Cambridge University Press.

\bibitem{ma2018}  Ma, Y.,  Gunatilaka, D.,  Li, B.,  Gonzalez, H., \&  Lu, C. (2018). Holistic cyber-physical management for dependable wireless control systems, {\it ACM Trans. Cyber-Phys. Syst., 3(1)}, Article 3.

\bibitem{Mesbahi} Mesbahi, M., \& Egerstedt, M. (2010). {\it Graph Theoretic Methods for Multiagent Networks}. Princeton University Press.

\bibitem{Molin}  Molin, A., \&  Hirche, S. (2014). Price-based adaptive scheduling in multi-loop control systems with resource constraints, {\it IEEE Trans. Autom. Control, 59(12)}, 3282--3295.

\bibitem{Park2018}  Park, P.,  Ergen, S. C.,  Fischione, C.,  Lu, C., \&  Johansson, K. H. (2018). Wireless network design for control systems: A survey, {\it IEEE Commun. Surv. Tutor., 20(2)}, 978--1013.

\bibitem{Pasqualetti2014}  Pasqualetti, F.,  Borra, D., \&  Bullo, F. (2014). Consensus networks over finite fields, {\it Automatica, 50},  349--358.

\bibitem{Pulikottil2021}  Pulikottil, T.,  Estrada-Jimenez, L.,  Rehman, H.,  Barata, J.,  Nikghadam-Hojjati, S., \&  Zarzycki, L. (2021). Multi-agent based manufacturing: Current trends and challenges, In {\it Proc. 26th IEEE International Conference on Emerging Technologies and Factory Automation}, Vasteras, Sweden.

\bibitem{Quevedo2013}  Quevedo, D. E.,  Ahl\'{e}n, A., \&  Johansson, K. H. (2013). State estimation over sensor networks with correlated wireless fading channels, {\it IEEE Trans. Autom. Control, 58(3)}, 581--593.

\bibitem{Quevedo2012}  Quevedo, D. E.,  Ahl\'{e}n, A.,  Leong, A. S., \&  Dey, S. (2012). On Kalman filtering over fading wireless channels with controlled transmission powers, {\it Automatica, 48(7)}, 1306--1316.

\bibitem{Sadeghi2008} Sadeghi, P.,  Kennedy, R. A.,  Rapajic, P. B.,  \& Shams, R.  (2008). Finite-state Markov modeling of fading channels-a survey of principles and applications,  {\it IEEE Signal Processing Magazine, 25(5)},  57--80. 

\bibitem{Sundaram2013}  Sundaram, S., \&  Hadjicostis, C. N. (2013). Structural controllability and observability of linear systems over finite fields with applications to multi-agent systems, {\it IEEE Trans. Autom. Control, 58(1)}, 60--73.

\bibitem{valerio2021} Valerio, P., (2021). Industry 4.0 requires massive IoT and seamless connectivity, IoT Times. [Online]. Available: https://iot.eetimes.com/industry-40-requires-massive-iot-and-seamless-connectivity

\bibitem{vitturi2019}  Vitturi, S.,  Zunino, C., \&  Sauter, T. (2019). Industrial communication systems and their future challenges: Next-generation Ethernet, IIoT, and 5G, {\it Proc. IEEE, 107(6)}, 944--961.

\bibitem{wangs2022} Wang, S.,  Li, P.,  Zhu, S., \&  Chen, C. (2022). Opportunistic wireless control over state-dependent fading channels, In {\it Proc. 61th IEEE Conference on Decision and Control}, Canc\'{u}n, Mexico (pp. 3896--3901).

\bibitem{wuy2019}  Wu, Y.,  Sun, X.-M.,  Zhao, X., \&  Shen, T. (2019). Optimal control of Boolean control networks with average cost: A policy iteration approach, {\it Automatica, 100},  378--387.

\bibitem{zhang2006}  Zhang, L., \&  Hristu-Varsakelis, D. (2006). Communication and control co-design for networked control systems, {\it Automatica, 42(6)}, 953--958.

\bibitem{zhangq1999}  Zhang, Q., \&  Kassam, S. (1999). Finite-state Markov model for Rayleigh fading channels, {\it IEEE Trans. Commun., 47(11)}, 1688--1692.

\bibitem{Zhao2011}  Zhao, Y.,  Li, Z., \&  Cheng, D. (2011). Optimal control of logical control networks, {\it IEEE Trans. Autom. Control, 56(8)},  1766--1776.


\bibitem{zhu2018}  Zhu, Q.,  Liu, Y.,  Lu, J., \&  Cao, J. (2018) On the optimal control of Boolean control networks, {\it SIAM J. Control Optim., 56(2)}, 1321--1341.
\end{thebibliography}

\end{document}